\documentclass[aps,prb,reprint,superscriptaddress]{revtex4-2}

\usepackage{dcolumn}
\usepackage{bm}
\usepackage{graphics,graphicx} 
\usepackage{amsmath,amssymb}
\usepackage{booktabs}
\usepackage{subcaption}
\usepackage[font=footnotesize]{caption}
\usepackage{romannum}
\usepackage{comment}
\usepackage{algorithm2e}
\usepackage{chemarrow}
\usepackage[none]{hyphenat}
\usepackage{xcolor}
\usepackage{multirow}
\usepackage{diagbox} 
\usepackage{array}   
\usepackage{tikz}
\usepackage{hyperref}
\usepackage{xr-hyper}
\usepackage{float}

\usepackage{textcase}  

\begin{abstract}
A population of individuals with the same genes can present heterogeneous traits (phenotypes).~The prevalence of this heterogeneity can be explained as a \textit{bet-hedging} strategy that improves the population proliferation rate (fitness) in fluctuating environments.~The phenotype distribution is influenced by factors such as competition between phenotypes, the duration of environmental states, and the rate of phenotype-switching.~We illustrate these effects in a system where both the environment and the phenotype can adopt two states.~This system includes scenarios such as symmetric bet-hedging and dormant-proliferating phenotypes.~We examine how environmental and phenotypic states share mutual information, measured in bits, and explore the relationship between this information and population fitness.~We propose that when fitness is measured relative to the case where phenotype and environment are independent, information and fitness can be treated as equivalent measures.~We investigate strategies that individuals can use to improve this information, such as adjusting the rates of proliferation and phenotype-switching relative to the environmental fluctuation rate.~Through these strategies, with fixed marginal distributions, an increase in information implies an increase in population fitness.~We also identify limits to the maximum achievable fitness and information and discuss the value of the information in terms of this new normalized fitness.~Our framework offers new insights into how organisms adapt to fluctuating environmental conditions.
\end{abstract}

\makeatletter
\newcommand*{\addFileDependency}[1]{
  \typeout{(#1)}
  \@addtofilelist{#1}
  \IfFileExists{#1}{}{\typeout{No file #1.}}
}
\makeatother

\newcommand*{\myexternaldocument}[1]{%
    \externaldocument{#1}%
    \addFileDependency{#1.tex}%
    \addFileDependency{#1.aux}%
}

\myexternaldocument{Supplement}

\begin{document}
\title{\LARGE \bf
Information and fitness in two-state systems: self-replicating individuals in a fluctuating environment.}


\preprint{APS/123-QED}
\author{Poulami Chatterjee } \thanks{These authors contributed equally to this work.}
\affiliation{Department of Electrical and Computer Engineering, Newark, Delaware, USA}
\author{C\'esar Nieto }\thanks{These authors contributed equally to this work.}
\affiliation{Department of Electrical and Computer Engineering, Newark, Delaware, USA}
\author{Juan Manuel Pedraza}
\affiliation{Department of Physics, Universidad de los Andes, Bogota, Colombia}
\author{Abhyudai Singh}
\affiliation{Department of Electrical and Computer Engineering, Biomedical Engineering, Mathematical Sciences, Center of Bioinformatics and Computational Biology, University of Delaware, Newark, DE, USA}

\maketitle



\begin{figure*}[ht!]
    \includegraphics[width=\linewidth]{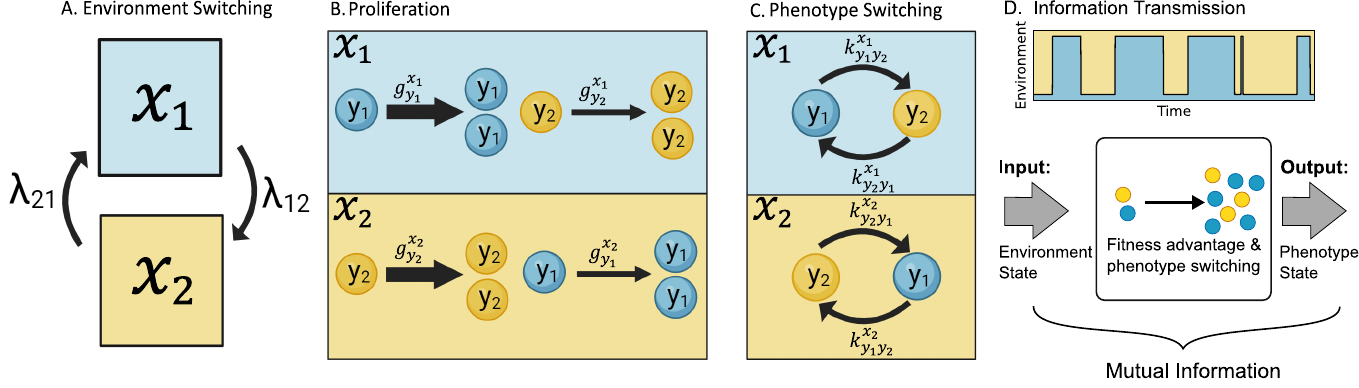}
    \caption{\textbf{Main variables of population dynamics in a fluctuating environment:} \textbf{A.} We consider two different environmental states, $x=x_1$ (blue) and $x=x_2$ (yellow).~The environment randomly switches (curved arrows) from $x_i$ to $x_j$ at rate $\lambda_{ij}$.~\textbf{B.~} The individuals (spheres) can have two phenotype states $y=y_1$ (blue) and $y=y_2$ (yellow).~Individuals with phenotype $y_j$ proliferate at a rate (represented by the straight arrow) $g^{x_i}_{y_j}$ in environment $x_i$.~The fastest growing phenotype (bold arrow) in environment $x_i$ is $y_i$ (They share similar colors).~\textbf{C.} Individuals perform random transitions (curved arrows) from phenotype $y_j$ to $y_{j'}$ at a rate $k^{x_i}_{y_jy_{j'}}$ in environment $x_i$ where $i,j,j'\in\{1,2\}$ and $j\neq j'$.~\textbf{D.} The environment follows a random process taking one of the two possible states.~In the context of information theory, given an environmental state the population will show a phenotype distribution.~Hence, the environment (input) and the phenotype state (output) share mutual information.}
    \label{fig:schematic_switching}
\end{figure*}

\section{Introduction}

Organisms are self-replicating entities whose fitness (proliferation rate) depends on their individual phenotypes (observable characteristics) and how they fit with environmental conditions~\cite{dekel2005optimality,aube2024genotype,xue2019environment,de2019exploration}.~However, environmental conditions in nature are inherently dynamic and change over time \cite{cvijovic2015fate}.~In these fluctuating environments, the fitness of the population can be improved when population uses a strategy known as \textit{bet-heging}~\cite{levien2021non,de2023effective,kussell2005phenotypic,thattai2004stochastic}.~A population using this strategy will be composed by diverse phenotypes even if they are not optimal for the current environmental state~\cite{fernandez2024evolution,morawska2022diversity}.~In this way, under sudden environmental change, among the alternative phenotypes, there is typically at least one that can survive and proliferate in the new environment, allowing the population to persist in the long term.~Depending on the environmental state, the phenotype with the highest proliferation rate is referred to as the fittest phenotype, while the one with the lowest proliferation rate is considered unfit.~Experimental evidence of bet-hedging includes the metabolism of carbon sources~\cite{hawkins2006regulatory,solopova2014bet}, lysogen-lytic phenotypes in the lambda phage~\cite{maslov2015well}, the synthesis of flagella in bacteria~\cite{bonifield2003flagellar}, adaptation of coral to maritime environmental conditions~\cite{kenkel2016gene}, heritable non-genetic phenotypes enriched for stress responses in microbes~\cite{grissom2024heritable}, bacterial persistence~\cite{balaban2004bacterial,patra2013population,hernandez2024plasmid,diamond2021buying, rahman2025rethinking}, rare cell variability driving drug resistance in cancer~\cite{shaffer2017rare}, drug-tolerant persisters in HER2+ breast cancer~\cite{chang2022ontogeny} and disruption of cellular memory affecting drug~resistance~\cite{harmange2023disrupting}.

There are multiple ways random alternative phenotypes can emerge in a isogenic population.~One of them is phenotypical switching, in which the cell phenotype can change over time~\cite{henrion2023fitness}.~This phenomenon is related to the random properties of gene expression~\cite{josephs2018determining,vahdat2024capturing,zhang2025stochastic}.~We say that the switching is responsive when the switching rate depends on the environmental state~\cite{siddiq2024plasticity,aube2024genotype}, and it is random when the switching rate is independent of the environment ~\cite{thattai2004stochastic,singh2023probing}.~Another mechanism that can alter the phenotype distribution is the effect of phenotype competition~\cite{mattingly2022collective}.~Since organisms are self-replicative, a phenotype with a higher proliferation rate will generate more copies of itself, increasing its proportion within the~population~\cite{scheiner1993genetics}.

The impact of environmental fluctuations on phenotypic distributions, as well as the ways in which populations respond to changing environments, remains an open question~\cite{aube2024genotype,bernhardt2020life}.~It is expected that the distribution of phenotypes stores information about the environment.~This is mainly, because for particular environmental states, fit phenotypes proliferate faster, changing the phenotype distribution.~Through this assumption, environment and population states can share mutual information~\cite{libby2007noisy,koonin2016meaning} as captured by the classical information theory~\cite{shannon2001mathematical,bermek2024form}.~The interplay between information and fitness has established important concepts such as the cost of information~\cite{donaldson2010fitness}, which gives some meaning to the information an individual can acquire from the environment.~In this way, by relating the information to an improvement in fitness, it is possible to impose limits on how accurate, in terms of information theory, a regulatory pathway must be to be functional from a biological perspective~\cite{hidalgo2014information}.

There are currently theoretical approaches where the mutual information between the environment and phenotype distributions has been estimated for ideal cases~\cite{pedraza2018noise,moffett2022minimal}.~We can include the case where environment and phenotypes change faster than the time scale of phenotype competition.~Therefore, environment and phenotype can be approximated as random variables with weak temporal autocorrelation~\cite{rivoire2011value,taylor2007information}.~In these contributions, they found that in particular cases information and fitness are equivalent quantities.~Our contribution aims to reach a more general equivalence between information and fitness, including properties such as the finite response time of phenotype switching, and the fitness advantage between environmental and individual states.

This article is structured as follows.~First, we model the dynamics of the fraction of a two-state phenotype.~We use a general approach that includes phenotype competition and switching in the presence of a two-state environment with memory-less transitions.~We demonstrate how average fractions and average environmental time duration can be used to define probability densities, allowing information to flow from the environment (input) to the phenotype (output) as an information channel.~We introduce the concept of normalized fitness, defined as the difference between the growth rate under responsive bet-hedging and the growth rate when the distributions of the phenotypes are independent of the environment, normalized by the maximum achievable relative growth rate.~With fixed environmental and phenotype probabilities, this normalized fitness is found to increase with the amount of information, establishing a \textit{one-to-one} relationship between these quantities.~This means that an increase in information implies an increase in the fitness of the population.~Furthermore, we find that the derivative of this fitness function with respect to information decreases as information increases.~Therefore, when population information is high, a marginal increase in information will result in a modest increase in fitness.~Finally, we examine the limitations of our approach through the relevant example of dormant-proliferating phenotypes, highlighting how our fitness metric compares to a system without constant marginal distributions.

\section{Model Formulation}

\subsection{Environment and population dynamics} 
We denote the environment as $x$ as a random process which can adopt two states $x_i$ and $x_j$, ${(i, j) \in \{(1, 2), (2,1)\}}$.~These states can represent multiple environmental conditions, including different temperatures, chemical composition, and nutrient availability~\cite{slein2023effects,vermeersch2022microbes}.~The environment fluctuates stochastically between the state ${x_i\rightarrow x_j}$ (${x_j\rightarrow x_i}$) with a constant rate $\lambda_{ij}$ ($\lambda_{ji}$) (Fig.~\ref{fig:schematic_switching}A).~Therefore, environmental switching occurs as a telegraph process, and the duration of the environmental state $x_i$ follows an exponential distribution with average $1/\lambda_{ij}$.~Individuals are described by their phenotype $y$ which can take two possible states designated as ${y_i \in \{y_1, y_2\}}$.~Depending on the environmental state $x_i$, each phenotype ${y_j}$ will have a a proliferation rate denoted as $g^{x_i}_{y_j}$ (Table \ref{tab:growth_payoff}).~The fittest phenotype $y_i$ in the environment $x_i$ is selected so that $g^{x_i}_{y_i} \geq g^{x_i}_{y_j}$ (Fig.~\ref{fig:schematic_switching}C).~

\begin{table}[h]

    \begin{minipage}[t]{0.45\linewidth}
        
        \renewcommand{\arraystretch}{1.6} 
        \begin{tabular}{|c|c|c|}
            \hline
            \multirow{2}{*}{\diagbox[width=4.5em, height=3.5em]{\textbf{State}}{\textbf{Env.}}} 
            & \multicolumn{1}{c|}{\textbf{$x_1$}} 
            & \multicolumn{1}{c|}{\textbf{$x_2$}} \\
            & & \\
            \hline
            \textbf{$y_1$} 
            & $g^{x_1}_{y_1}$ 
            & $g^{x_2}_{y_1}$ \\
            \hline
            \textbf{$y_2$} 
            & $g^{x_1}_{y_2}$ 
            & $g^{x_2}_{y_2}$ \\
            \hline
        \end{tabular}
        \caption{Fitness matrix in environment $x_i$ for the individuals with phenotype $y_j$ in the general case.~}
        \label{tab:growth_payoff}
    \end{minipage}
    \hspace{0.05\linewidth} 
    \begin{minipage}[t]{0.45\linewidth}
        
        \renewcommand{\arraystretch}{1.6} 
        \begin{tabular}{|c|c|c|}
            \hline
            \multirow{2}{*}{\diagbox[width=4.5em, height=3.5em]{\textbf{State}}{\textbf{Env.}}} 
            & \multicolumn{1}{c|}{\textbf{$x_1$}} 
            & \multicolumn{1}{c|}{\textbf{$x_2$}} \\
            & & \\
            \hline
            \textbf{$y_1 \rightarrow y_2$} 
            & $k^{x_1}_{y_1y_2}$
            & $k^{x_2}_{y_1y_2}$ \\
            \hline
            \textbf{$y_2 \rightarrow y_1$} 
            & $k^{x_1}_{y_2y_1}$
            & $k^{x_2}_{y_2y_1}$ \\
            \hline
        \end{tabular}
        \caption{Phenotypical switching rates in the general case.}
        \label{tab:switching}
    \end{minipage}
\end{table}

To implement the bet-hedging strategy, individuals must perform stochastic transitions between their phenotypes (Fig.~\ref{fig:schematic_switching}C).~These transitions are also known as phenotypical switching, since they are approximated to occur spontaneously with no delays.~The switching rate from phenotype $y_j$ to $y_k$ in the environment $x_i$ is indicated as $k^{x_i}_{y_jy_k}$, (Table~\ref{tab:switching}).~We consider that the switching is responsive when it depends on the environment; otherwise, the switching is non-responsive or random.~When the rate of transition from the least fit to the fittest phenotype is greater ${k^{x_i}_{y_jy_i}>k^{x_i}_{y_iy_j}}$, the phenotypical switch gives advantage to the population.

The probability of finding an individual with the phenotype $y_j$ will be associated with the fraction $f_{y_j}$ that corresponds to the number of individuals with the phenotype $y_j$ over the total number of individuals in the population.~By neglecting random fluctuations in population growth, the fraction $f_{y_j}$ can be modeled using deterministic differential equations.~The general dynamics for $f_{y_j}$ in the environment $x_i$ is as follows~\cite{skanata2016evolutionary,davila2024extinction} (see supplementary information S1 for details):
\begin{align}\label{general_model}
\frac{df_{y_j}}{dt} =
    k^{x_i}_{y_k y_j} (1 - f_{y_j}) - k^{x_i}_{y_j y_k} f_{y_j} \nonumber\\
    +(g^{x_i}_{y_j} - g^{x_i}_{y_k}) f_{y_j} (1 - f_{y_j}) 
\end{align}
where $j\neq k$.~For example, the dynamics of $f_{y_1}$ in different environments follows:
\begin{equation}\label{fraction_y1}
\frac{df_{y_1}}{dt} =
\begin{cases} 
    k^{x_1}_{y_2y_1} (1 - f_{y_1}) - k^{x_1}_{y_1y_2} f_{y_1} \\
    +(g^{x_1}_{y_1} - g^{x_1}_{y_2}) f_{y_1} (1 - f_{y_1}), & \text{if } x = x_1, \\[10pt]
    
    k^{x_2}_{y_2y_1} (1 - f_{y_1}) - k^{x_2}_{y_1y_2} f_{y_1} \\
    +(g^{x_1}_{y_2} - g^{x_2}_{y_2}) f_{y_1} (1 - f_{y_1}), & \text{if } x = x_2.
\end{cases}
\end{equation}
During an environmental switch, $f_{y_1}$ follows continuous boundary conditions.~This means that the value of $f_{y_1}$ remains unchanged immediately before and after the~switch.

Using the dynamics of $f_{y_j}$ given $x_i$, we establish a relationship between the phenotypical state and the environment.~The quality of this link can be quantified using information theory.~In this framework, the environmental state acts as an input and the phenotypical dynamics represent the response to that environmental perturbation transmitting the information to the resulting phenotypic states (the output), as illustrated in Fig.~\ref{fig:schematic_switching}D.~With these ideas, in the next section we will explore concrete examples of environmental switching and phenotype fitness.


\subsection{Single-individual limit}

A relevant and simple limiting case arises when the phenotype confers no selective advantage or the timescales are not long enough for the phenotype competition.~This means that the growth rate can take as uniform ($g^{x_i}_{y_j}=g$) across all environments ($x_i$) and phenotypes ($y_j$).~In this scenario, the difference in growth rates between any two phenotypes in any environment becomes zero ($g^{x_i}_{y_j} - g^{x_i}_{y_k}=0$), simplifying the general model described by equation~\eqref{general_model} into a linear system solvable with standard matrix methods~\cite{shin2010linear}.~Although not extensively explored in this article, this linear limit holds particular relevance in several contexts.~Firstly, it can approximate biosensor systems in which the engineered gene circuit imposes a minimal metabolic burden on host cells~\cite{van2010microbiology}.~Secondly, it is applicable to experimental setups where the cell offspring is discarded, effectively eliminating proliferation advantages between phenotypes, as seen in cell microfluidic devices such as the \textit{mother machine}~\cite{taheri2015cell,duran2020slipstreaming}.~Finally, the single-individual limit becomes relevant when environmental fluctuations and phenotypical switching occur on timescales much faster than the proliferation rate.~In such cases, an individual can pass through multiple phenotypes during its lifespan in response to environmental changes, with survival ability being the primary selective pressure rather the long-term proliferation rate~\cite{moffett2022minimal}.~Compared to models that incorporate fitness differences, these single-individual scenarios generally predict relatively lower levels of mutual information between the environment and the phenotypical state.

\subsection{Model of symmetric bet-hedging}\label{thattai}
We can simplify our generalized model \eqref{general_model} according to the structure proposed by Thattai \textit{et.~al.}~\cite{thattai2004stochastic}.~This model is one of the simplest and is usually used for explaining the concept of bet-hedging in fluctuating environments.~In this case, both the growth and the switching rate matrix are symmetric under the change of label $1\rightarrow2$.~Specifically, for growth rates, we have $g^{x_1}_{y_2}=g^{x_2}_{y_1}=\rho$ and $g^{x_1}_{y_1}=g^{x_2}_{y_2}=\mu$ with $\mu>\rho$ (Table \ref{tab:tatai_growth_payoff}).~The phenotype switching rates, considered as responsive, can be written as $k^{x_1}_{y_1y_2} = k^{x_2}_{y_2y_1}=k_a$ and $k^{x_1}_{y_2y_1} = k^{x_2}_{y_1y_2}=k_b$, with $k_b\geq k_a$ (Table \ref{tab:tatai_switching}).
\begin{table}[ht!]
    
    \begin{minipage}{0.45\linewidth}
        
        \renewcommand{\arraystretch}{1.6} 
        \begin{tabular}{|c|c|c|}
            \hline
            \multirow{2}{*}{\diagbox[width=4.5em, height=3.5em]{\textbf{State}}{\textbf{Env.}}} 
            & \multicolumn{1}{c|}{\textbf{$x_1$}} 
            & \multicolumn{1}{c|}{\textbf{$x_2$}} \\
            & & \\
            \hline
            \textbf{$y_1$} 
            & $\mu$ 
            & $\rho$ \\
            \hline
            \textbf{$y_2$} 
            & $\rho$ 
            & $\mu$ \\
            \hline
        \end{tabular}
        \caption{Fitness matrix for the symmetric fitness model ($\mu>\rho$).}
        \label{tab:tatai_growth_payoff}
    \end{minipage}
    \hspace{0.05\linewidth} 
    \begin{minipage}{0.45\linewidth}
        
        \renewcommand{\arraystretch}{1.6} 
        \begin{tabular}{|c|c|c|}
            \hline
            \multirow{2}{*}{\diagbox[width=5em, height=3.5em]{\textbf{State}}{\textbf{Env.}}} 
            & \multicolumn{1}{c|}{\textbf{$x_1$}} 
            & \multicolumn{1}{c|}{\textbf{$x_2$}} \\
            & & \\
            \hline
            \textbf{$y_1 \rightarrow y_2$} 
            & $k_{a}$ 
            & $k_{b}$ \\
            \hline
            \textbf{$y_2 \rightarrow y_1$} 
            & $k_{b}$ 
            & $k_{a}$ \\
            \hline
        \end{tabular}
        \caption{Switching Rates for the symmetric fitness model ($k_{b}>k_{a}$).}
        \label{tab:tatai_switching}
    \end{minipage}
\end{table}
\subsection{Model of dormant-proliferating phenotypes}\label{persistent}
Another relevant two-state system is a population of individuals taking dormant-proliferating phenotypes.~This scenario can describe cases such as persister versus proliferating cells under antibiotic stress~\cite{patra2013population, guha2025priestia, hossain2023escherichia, bokes2025optimisation} and non-mobile versus proliferating alga in dry-wet environments~\cite{tang2024bet}.~The environment $x_1$ can be interpreted as the stressful scenario (antibiotic or dry environment) while the environment $x_2$ represents regular growth conditions.~Individuals can switch between a dormant phenotype, which has a negligible growth rate, and a proliferating phenotype, which grows rapidly under regular conditions but exhibits a negative growth rate under stressful conditions, implying individual death.~Mathematically, the rates can be represented as: $ {g^{x_1}_{y_2}=g^{x_2}_{y_2}=0}$, ${g^{x_1}_{y_1}=-\mu_1}$ and ${g^{x_2}_{y_1}=\mu_2}$ (Table \ref{tab:persistence_growth_payoff}).~The particular values of the switching rates (Table \ref{tab:persistence_switching}) are not very relevant; our goal is simply to establish the bet-hedging focusing on the conceptual properties.~Therefore, we will consider the switching rates to have the same value: $k^{x_1}_{y_1y_2} = k^{x_2}_{y_2y_1}=k^{x_1}_{y_2y_1} = k^{x_2}_{y_1y_2}=k$.
\begin{table}[ht!]
    
    \begin{minipage}{0.45\linewidth}
        
        \renewcommand{\arraystretch}{1.6} 
        \begin{tabular}{|c|c|c|}
            \hline
            \multirow{2}{*}{\diagbox[width=4.5em, height=3.5em]{\textbf{State}}{\textbf{Env.}}} 
            & \multicolumn{1}{c|}{\textbf{$x_1$}} 
            & \multicolumn{1}{c|}{\textbf{$x_2$}} \\
            & & \\
            \hline
            \textbf{$y_1$} 
            & $0$ 
            & $0$ \\
            \hline
            \textbf{$y_2$} 
            & $-\mu_1$ 
            & $\mu_2$ \\
            \hline
        \end{tabular}
        \caption{Fitness matrix for the model of dormant-proliferating phenotypes.}
        \label{tab:persistence_growth_payoff}
    \end{minipage}
    \hspace{0.05\linewidth} 
    \begin{minipage}{0.45\linewidth}
        
        \renewcommand{\arraystretch}{1.6} 
        \begin{tabular}{|c|c|c|}
            \hline
            \multirow{2}{*}{\diagbox[width=5em, height=3.5em]{\textbf{State}}{\textbf{Env.}}} 
            & \multicolumn{1}{c|}{\textbf{$x_1$}} 
            & \multicolumn{1}{c|}{\textbf{$x_2$}} \\
            & & \\
            \hline
            \textbf{$y_1 \rightarrow y_2$} 
            & $k$ 
            & $k$ \\
            \hline
            \textbf{$y_2 \rightarrow y_1$} 
            & $k$ 
            & $k$ \\
            \hline
        \end{tabular}
        \caption{Switching Rates for the model of dormant-proliferating phenotypes.}
        \label{tab:persistence_switching}
    \end{minipage}
\end{table}

\section{Environment and Population Statistics}
\begin{figure}[ht!]
\includegraphics[width=\linewidth] {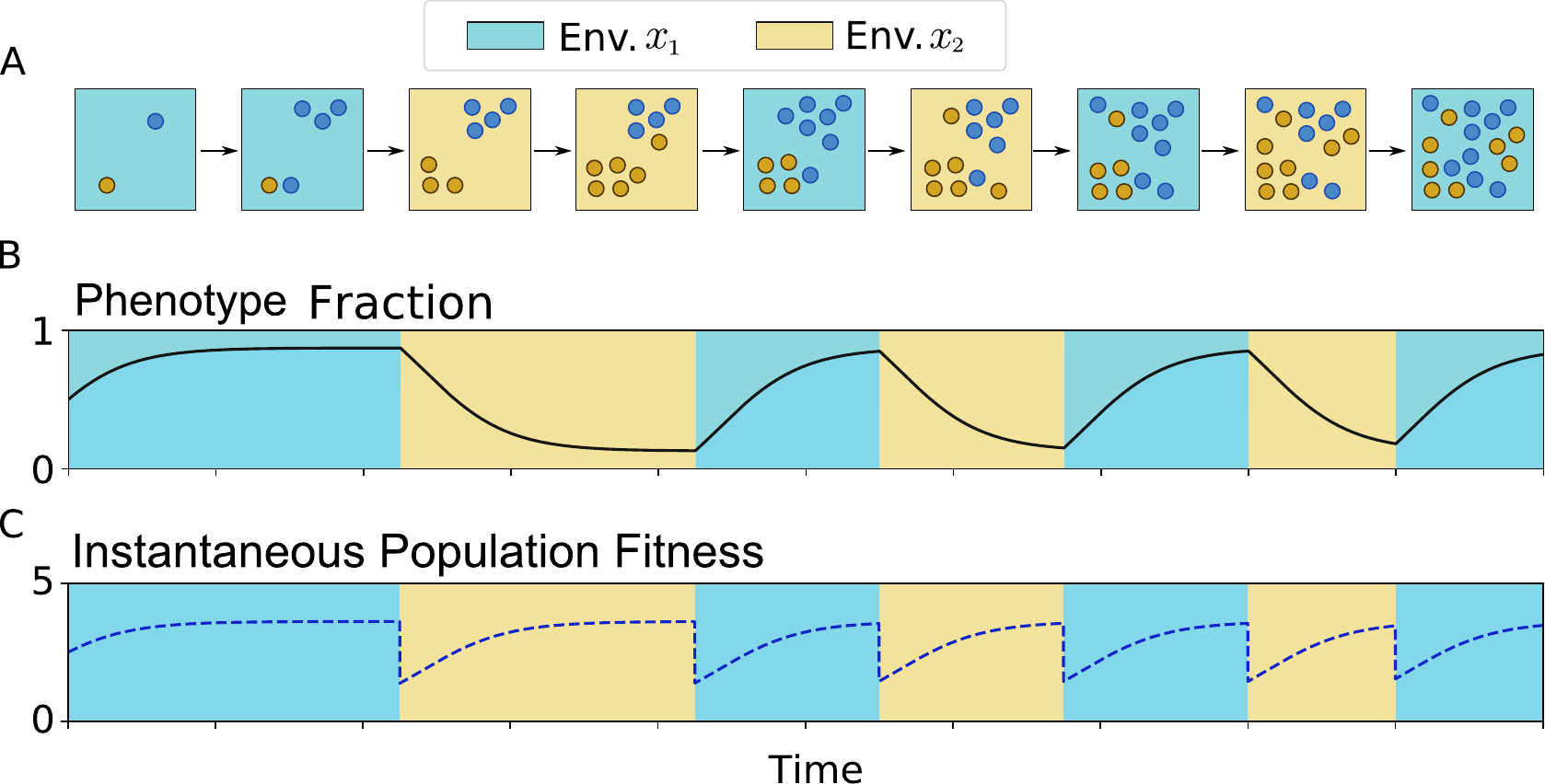}
    \caption{\textbf{Phenotype distribution over time in a two-state fluctuating environment and its fitness properties:} \textbf{A.}~Schematic representation of the population dynamics.~Individuals proliferate and switch their phenotype depending on the environment.~\textbf{B.} As a result of the fluctuating environment, the fraction of individuals in each phenotype evolves over time.~The fraction of individuals with phenotype $y_1$ is derived from the eq.\eqref{general_model}.~The fraction will be related to the probability of finding an individual in that phenotype.~\textbf{C.} The population fitness is defined as the average growth rate over all the instantaneous phenotype fractions.}
    \label{fig:switching_fraction_compact}
\end{figure}

A schematic representation of how individuals proliferate and change in a fluctuating environment is presented in Fig.~\ref{fig:switching_fraction_compact}A in which each different environment is represented by a color (blue or yellow).~The dynamics of the population fraction in response to the environment switch is shown in Fig.~\ref{fig:switching_fraction_compact}B.~Observe that if the environment is fixed, equation \eqref{general_model} predicts a steady fraction where the fittest phenotype is more abundant than the other one.~Fig.~\ref{fig:switching_fraction_compact}B shows how, after an environmental switch, $f_{y_i}$ needs some time to reach a new steady value.

At a particular time, the probability of finding an individual with the phenotype $y_i$ is related to the fraction $f_{y_i}$.~With knowledge of this fraction, it is possible to estimate some dynamics quantities.~For example, population fitness $\gamma$ is defined as the average growth rate over the phenotype distribution:
\begin{align}\label{eq:fgr}
    \gamma(t)\triangleq\sum_{y_j}g^{x_i(t)}_{y_1}f_{y_j}(t).
\end{align}
We present an example of the dynamics of $\gamma$ over time in Fig.~\ref{fig:switching_fraction_compact}C.~Notably, after an environmental switch, most of the population is in the least fit state, which makes $\gamma(t)$ lower.~As the population relaxes to the new fittest state, the population fitness increases.

An important limitation of expression \eqref{eq:fgr} is that the mean growth rate depends on the particular environmental trajectory.~To have a more general approach, we will study the transmission of information in a time-independent way.~To achieve this, we will estimate the general probabilities $P_{x_i}$ for  the environmental state and $P_{y_i}$ for the phenotypical state.~The environmental probability can be estimated from the dynamics of $x$ as:
\begin{equation}\label{eq:pxi}
    P_{x_i}=\lim_{T\rightarrow \infty} \frac{1}{T} \int_{0}^{T} \mathbf{1}_{x_i}(x) \, dt.
\end{equation}
where $\mathbf{1}_{x_i}$ is the indicator function
\[
\mathbf{1}_{x_i}(x) =
\begin{cases} 
    1, & \text{if } x = x_i, \\
    0, & \text{if } x \neq x_i.
\end{cases}
\]
This probability is equal to the well-known expression for a telegraph process,
$P_{x_i}=\frac{\lambda_{ji}}{\lambda_{ji}+\lambda_{ij}}$~\cite{gardiner2009stochastic}.
To define the probability $P_{y_i}$ of finding an individual in the phenotype $y_j$ at any time, we follow a previous approach~\cite{skanata2016evolutionary} in which $P_{y_j}$ is associated with the time-averaged fraction of individuals in the phenotype $y_j$ as:
\begin{equation}\label{eq:pyi}
    P_{y_i}\triangleq\langle f_{y_j} \rangle=\lim_{ T\rightarrow \infty} \frac{1}{T}\int_{0}^{T} f_{y_j}(t) \, dt,
\end{equation}
The joint probability $P_{x_iy_j}$ is estimated using the fraction and the indicator function as follows:
\begin{equation}\label{eq:pxiyi}
     P_{x_i y_j} \triangleq \lim_{T\rightarrow \infty} \frac{1}{T} \int_{0}^{T} f_{y_j}(t)
     \mathbf{1}_{x_i}(x)  dt
\end{equation}

This expression follows the properties of a probability distribution.~Specifically, we have 
\begin{equation}
    \sum_i\sum_j P_{x_iy_j}=1,
\end{equation}
which can be proven using $\sum_{j}f_{y_j}^{x_i}=1$.~Furthermore, the marginal distribution for the state of the phenotype $y_j$ follows $P_{y_j}=\sum_{x_i}P_{x_i,y_j}$ (see table \ref{tab:Probability Matrix}).~Similarly, the marginal distribution of the environmental state $x_i$ follows $P_{x_i}=\sum_{y_j}P_{x_i,y_j}$.
\begin{table}[ht!]
    
    \includegraphics[trim=7cm 21.2cm 7cm 4cm, clip, width=0.8\linewidth]{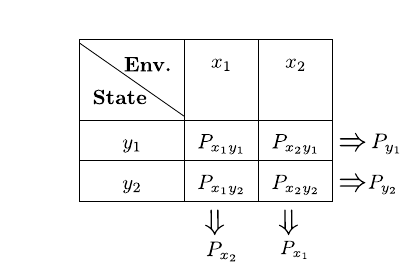} 
    \caption{Probability Matrix: The probability $P_{x_iy_j}$ of finding the environment $x_i$ and phenotype state $y_j$ and the marginal probabilities $P_{x_i}$ and $P_{y_j}$}
    \label{tab:Probability Matrix}
\end{table}

Particularly, in our simulations, we will use equations \eqref{eq:pxi}, \eqref{eq:pyi}, and \eqref{eq:pxiyi} to estimate these probabilities (See Supplementary Information, Algorithm 1 and Algorithm 2).~For the theoretical discussion, we will assume that these probabilities exist, and using their properties, it is possible to estimate some boundaries on the information and fitness.~As a starting point, we define the time-averaged fitness $\langle \gamma\rangle$ as the time-average of the expression \eqref{eq:fgr}, in terms of the probabilities $P_{x_iy_j}$ and the growth rate of each phenotype in each environmental state (Table \ref{tab:growth_payoff})
\begin{align}\label{gamma_prob}
    \langle \gamma\rangle \triangleq g^{x_1}_{y_1} P_{x_1y_1} +g^{x_1}_{y_2}P_{{x_1}y_2}
     +g^{x_2}_{y_1} P_{{x_2}y_1} + g^{x_2}_{y_2} P_{{x_2}y_2}.
\end{align}
With these probabilities, we can also define the mutual information between the environment and the phenotype distribution as follows:
\begin{align}\label{info}
I&\triangleq{\sum_{i,j=1}^{2} P_{x_iy_j} \log_2\left(\frac{P_{x_iy_j}}{P_{x_i}P_{y_j}}\right)}.
\end{align}

The mutual information quantifies the degree of knowledge about the input of the system (environment)  that can be gained by observing its output (phenotypical state) and vice-versa.~Once the definition of $I$ in~\eqref{info} includes the logarithm in base 2, this information is measured in units of bits.~Therefore, if the system shares mutual information $I$, the states that input and output share can be classified into $2^I$ generalized independent states.~For example if it is possible to classify unequivocally the state of the input between 8 possible states, a binary system shares information of 3 bit, ($\log_2(8)=3$ bit).~In this two-state environment, the maximum available information corresponds to the environmental entropy ${H(x)=-\sum_i P_{x_i}\log_2(P_{x_i})}$, which has a value of 1 bit when the probability of each environment is 0.5.

\begin{figure*}
    \includegraphics[width=0.85\linewidth]{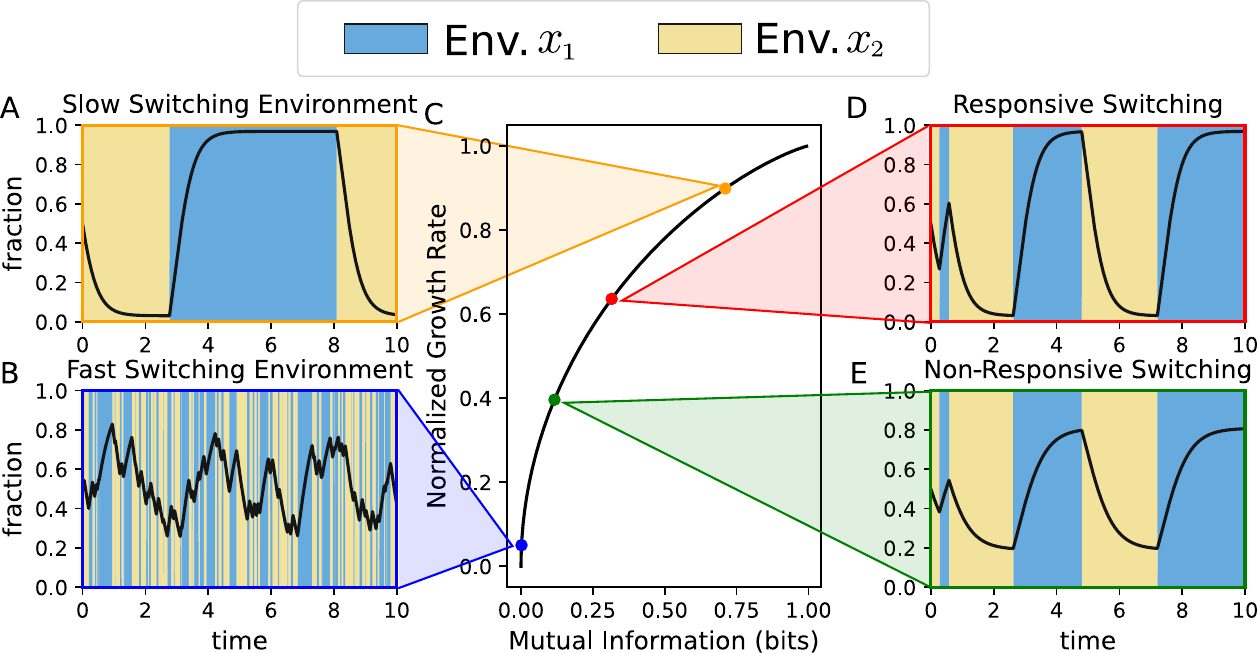}
    \caption{\textbf{Population fitness vs the mutual information in the symmetric-fitness model with four descriptive examples:} Depending on the particular population parameters (growth rate $g_{y_j}^{x_i}$ and phenotype switching rates $k_{y_jy_k}^{x_i}$) phenotype fractions will evolve differently.~Four examples with same relative growth rate ($g_{x_1}=g_{x_2}=\mu-\rho=3$): \textbf{A.} The environment changes with slower timescale than population (proliferation rate).~(Parameters: $\lambda_{12}=\lambda_{21}=0.1$, $k_a=0.1$, $k_b=1.2$).~\textbf{B.} Environmental switch occurs faster than population growth.~(Parameters: $\lambda_{12}=\lambda_{21}=10$, $k_a=0.1$, $k_b=1.2$).~\textbf{C.} Normalized growth rate following equation \eqref{eq:Gamma} related to mutual information using \eqref{info}.~Different dots have different colors representing each example.~\textbf{D.} Individuals switch their phenotype with rates dependent on the environment (responsive switching) ($k_a=0.1$, $k_b=1.2$) \textbf{E.} Individuals switch their phenotype with rates independent of the environment (non-responsive switching) ($k_a=0.5$, $k_b=0.5$).~Environment is considered to have a symmetric distribution with the same switching rates $\lambda_{12}=\lambda_{21}$.~Regardless the particular growth and switching rates, normalized fitness and mutual information will follow same relationship given by the expression \eqref{eq:gamaIsym} (black line in C).~In (A, B, D, E), the fraction of phenotype \( y_1 \) in different environments, denoted as \( f_{y_1} \), is represented by black lines, while the different environments are distinguished using two different colors: Blue for $x=x_1$ and yellow for $x=x_2$.}
    \label{fig:information}
\end{figure*}

\section{Information and fitness in simple particular cases}

An important limitation of the definition of average fitness $\langle \gamma\rangle$ in \eqref{gamma_prob} is that its value will depend on the particular values of $g^{x_i}_{y_j}$.~To study a robust metric, we define the fitness of the reference population $\langle\gamma\rangle_{ind}$ as the fitness of the average population when the environment and the phenotype are independent of each other, that is, when ${P_{x_iy_j}=P_{x_i}P_{y_j}}$.~More specifically, $\langle \gamma\rangle_{ind}$ is given by: 
\begin{align} \label{gamma_id}
    \langle\gamma\rangle_{ind}=&g^{x_1}_{y_1} P_{x_1} P_{y_1} +g^{x_1}_{y_2}P_{x_1}P_{y_2}
     +g^{x_2}_{y_1} P_{x_2}P_{y_1} + g^{x_2}_{y_2} P_{x_2}P_{y_2}.
\end{align}
We study the difference between $\langle\gamma\rangle$ and $\langle \gamma\rangle_{ind}$.~It is possible to demonstrate that this difference can be written as (see Supplementary information S2 for details)
\begin{align}\label{eq:relgr}
    \langle\gamma\rangle-\langle\gamma\rangle_{ind}=&g_{x_1}(P_{x_1y_1}-P_{x_1} P_{y_1})+g_{x_2}(P_{x_2y_2}-P_{x_2}P_{y_2}),
    \end{align}
where ${g_{x_1}=g_{y_1}^{x_1}-g_{y_2}^{x_1}>0}$ and ${g_{x_2}=g_{y_2}^{x_2}-g_{y_1}^{x_2}>0}$ are also known as the relative growth rate in each environment.~In addition, we will study the cases in which ${P_{x_iy_i}>P_{x_i}P_{y_i}}$.~This means that the probability of having the fittest phenotype in its respective environment is higher than when the environment and phenotype are independent.~

The highest achievable growth rate $\langle\gamma\rangle_{max}$ occurs when, for the environmental state $x_i$, all individuals have the phenotype $y_i$.~This results in the distribution $P_{x_iy_i}=P_{y_i}=P_{x_i}$, $i,j=1,2$ and the non-diagonal probabilities being 0.~After replacing this case in \eqref{eq:relgr}, it results in a relative fitness:
\begin{align}
    \langle\gamma\rangle_{max}-\langle\gamma\rangle_{ind}=&g_{x_1}(P_{x_1}-P^2_{x_1})+g_{x_2}(P_{x_2}-P^2_{x_2})\nonumber\\
    =&(g_{x_1} + g_{x_2})  P_{x_1}  (1 - P_{x_1}),
\end{align}
In which we used $P_{x_2}=1-P_{x_1}$.~In this way, we can define the normalized growth rate $\Gamma$ as the mean growth rate relative to the growth rate when the phenotype is independent of the environment over the highest achievable relative growth rate.~This normalized growth rate will be our measure of fitness and follows the formula: \begin{align}\label{eq:Gamma}
    \Gamma &= \frac{\langle \gamma\rangle-\langle\gamma\rangle_{ind}}{\langle\gamma\rangle_{max}-\langle\gamma\rangle_{ind}} \nonumber\\
    &= \frac{g_{x_1}  (P_{x_1y_1} - P_{x_1}  P_{y_1}) + g_{x_2}(P_{x_2y_2} - P_{x_2}  P_{y_2})}{ (g_{x_1} + g_{x_2})  P_{x_1}  (1 - P_{x_1})}.
\end{align}

\subsection{Information and fitness in the symmetric-fitness model}

To illustrate the relationship between mutual information $I$ and normalized fitness $\Gamma$, we will now apply the previous concepts to the symmetric fitness model in Section \ref{thattai}.~First, we consider that the environment has symmetric switching rates $\lambda_{12}=\lambda_{21}$.~This symmetric environment results in $P_{x_1}=P_{x_2}=0.5$.~The additional symmetry on switching rates let us conclude that the phenotype distribution follows $P_{y_1}=P_{y_2}=0.5$.~In this way, the distribution can be written, for instance, in terms of $P_{x1y1}$:
\begin{subequations}
    \begin{align}
        P_{x_1y_2}&=P_{x_1}-P_{x_1y_1}=\frac{1}{2}-P_{x_1y_1}\\
        P_{x_2y_1}&=P_{y_1}-P_{x_1y_1}=\frac{1}{2}-P_{x_1y_1}\\
        P_{x_2y_2}&=P_{x_2}-P_{x_2y_1}=P_{x_1y_1},
    \end{align}
\end{subequations}
and therefore, the fitness $\Gamma_{sym}$ and the information $I_{sym}$ for the symmetric-fitness model, with the subscript ${sym}$ standing for \textit{symmetric}, can be parameterized solely by the particular value of $P_{x_1y_1}$ which depends on the growth and switching rates.~These metrics can be simplified to:
\begin{subequations}
\begin{align}\label{eq:gamaIsym}
    \Gamma_{sym}=&4P_{x_1y_1}-1\\
    I_{sym}=&1+\left(1-2P_{x_1y_1}\right)\log_2\left(1-2P_{x_1y_1}\right)\nonumber\\
    &+2P_{x_1y_1}\log_2\left(2P_{x_1y_1}\right),
\end{align}
\end{subequations}
with the additional constrain $P_{x_1y_1}\in\{0.25,0.5\}$, which corresponds to the cases in which $P_{x_1y_1}>P_{x_1}P_{y_1}$.

In Fig.~\ref{fig:information} (black line in the central plot), we present the resultant relationship between $\Gamma_{sym}$ and $I_{sym}$ parameterized by $P_{x_1y_1}$.~The principal property of this relationship, which is in agreement with other simpler models~\cite{rivoire2011value,taylor2007information}, is that the growth rate \textit{is a monotonically increasing function of information}.~This means that every change in the rates at which mutual information increases results in an increase in the fitness of the population.

To obtain $\Gamma$ and $I$ for specific growth and switching rates, we simulated multiple fluctuating environments.~The joint distribution $P_{xy}$ is estimated from the integration of the fraction dynamics obtained from equations \eqref{eq:pyi}, \eqref{eq:pxiyi} and \eqref{eq:pxi}.~The probabilities were used to determine the values of $\Gamma$ and $I$ for each specific set of switching and growth rates.


When analyzing how the probability distribution depends on the particular rates, we can exemplify four relevant scenarios that are shown in Fig.~\ref{fig:information}.~They are the~following:

\begin{figure*}[ht!]
        \includegraphics[width=\linewidth]{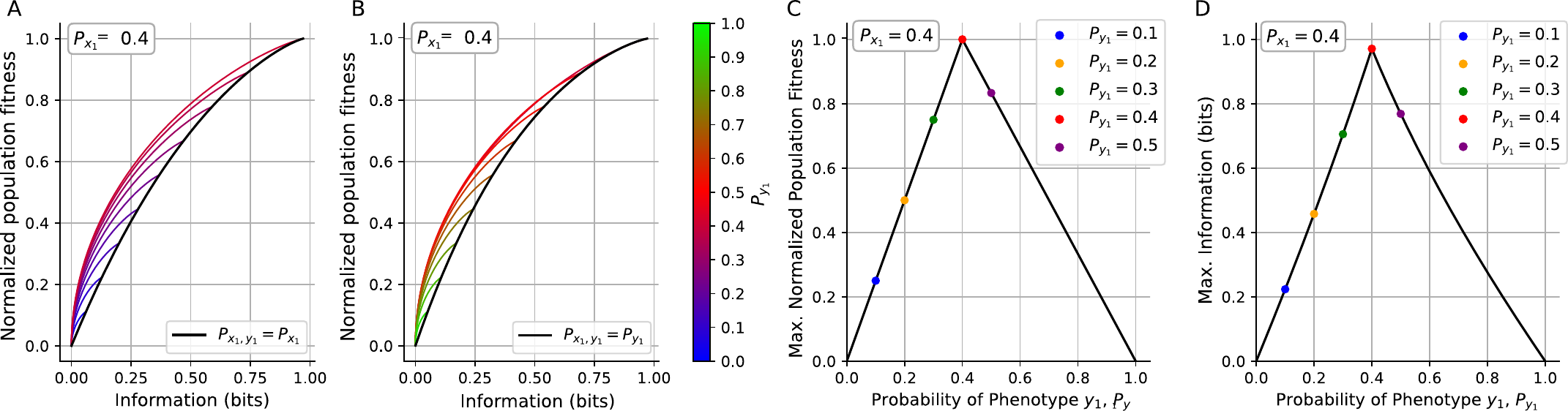}
    \caption{\textbf{Information and normalized fitness in the general case.~When $P_{x_1}$ and $P_{y_1}$ are fixed, increasing $I$ by increasing $P_{x_1y_1}$ results on increase in $\Gamma$.~The maximum achievable $\Gamma$ and $I$ are reached when $P_{x_1}=P_{y_1}$} \textbf{A.}  Normalized population fitness vs information by fixing $P_{x_1}=0.4$ and different values of $P_{y_1}\in(0,P_{x_1})$.~Without a specific mechanism (changing the phenotype switching or proliferation rate), probability $P_{x1y1}$ can increase within the interval $P_{x_1}P_{y_1}\leq P_{x_1y_1}\leq \min(P_{x_1},P_{y_1})$.~Each color corresponds to one specific value of $P_{y_1}$ (depicted in the color bar).~Black line represents the maximum achievable $\Gamma$ and $I$ $P_{x_1y_1}=\min(P_{x_1},P_{y_1})$.~\textbf{B.} Same as (A.) but for $P_{y_1}\in(P_{x_1},1)$~\textbf{C.} Dependence of the maximum $\Gamma$ with $P_{y_1}$.~The dots represent the maximum $\Gamma$ depicted in the vertical axis of (A,B).~\textbf{D.} Maximum achievable information vs $P_{y_1}$.~The dots represent the maximum $I$ depicted in the horizontal axis of (A,B).~Notably, $\Gamma$ and $I$, show a global maximum when $P_{y_1}=P_{x_1}$.}
    \label{fig:generalized}
\end{figure*}

\begin{itemize}
    \item \textit{Slow-switching environment}: In this scenario, the fluctuation and phenotypic-switching rates are faster than the environment transitions (Fig.~\ref{fig:information}A).~This relatively slow environmental dynamics enables the fittest phenotype to win the competition with the other phenotype in the corresponding environment, reaching the steady fraction over a relatively long period of time.~As a result, this situation provides a relatively high average growth rate and mutual information.
    
    \item \textit{Fast-switching environment}: In Fig.~\ref{fig:information}B the environmental switching rate is fast relative to the phenotype timescale.~In this rapid-environment scenario, the population does not remain in any given environment long enough to reach the equilibrium distribution in each environment.~This scenario results in low normalized fitness and low mutual information.

    \item \textit{Responsive phenotype switching:} In this case, phenotype switching favors the fittest phenotype (${k^{x_i}_{y_jy_i}>k^{x_i}_{y_iy_j}}$).~As presented in Fig.~\ref{fig:information}D for a sufficiently slow environment, the fraction of individuals in the least-fit phenotype is small and therefore the system will transmit important information, but not as high as the slow-switching environment.
    
    \item\textit{Non-responsive phenotype switching}:
In this case, the phenotypic switching rate does not favor the highest fit phenotype (${k^{x_i}_{y_jy_i}\leq k^{x_i}_{y_iy_j}}$).~As observed in Fig.~\ref{fig:information}E, the fraction of the least-fit phenotype is relatively high.~This effect reduces mutual information relative to the responsive switching case and therefore the average fitness of the population.
\end{itemize}

Fig.~\ref{fig:information}C shows how the normalized population fitness and mutual information for these four examples align with the general formula given by the expression \eqref{eq:gamaIsym}.

\section{Information and fitness in the general case}

This section examines the characteristics of the universal case, where the dynamics of the phenotypic and environment can have any arbitrary value.~In this case, the phenotypic distribution does not necessarily match the environmental distribution, i.e., $P_{x_i}\neq P_{y_i}$.~To analyze this scenario, we consider particular pairs of input and output distributions $P_{x_1}$ and $P_{y_1}$, respectively.~Without specific details of growth and switching rates, the joint distribution of environment-phenotype can be parameterized by the probability $P_{x_1y_1}$, which satisfies 
\begin{equation}\label{eq:P_{x_1}y1}
P_{x_1}P_{y_1}\leq P_{x_1y_1}\leq \min(P_{x_1},P_{y_1}).
\end{equation} 
The other probabilities of the joint distribution can be written in terms of the three parameters $P_{x_1},P_{y_1},P_{x_1y_1}$, obtaining the expression:
\begin{subequations}\label{eq:P_{x_1}y1n}
    \begin{align}
        P_{x_2y_1}&=P_{y_1}-P_{x_1y_1}\\
        P_{x_1y_2}&=P_{x_1}-P_{x_1y_1}\\
        P_{x_2y_2} &= 1-P_{x_2y_1}-P_{x_1y_2}-P_{x_1y_1} \nonumber\\
         &=1-P_{x_1}-P_{y_1}+P_{x_1y_1}.
    \end{align}
\end{subequations}
 Using these constraints \eqref{eq:P_{x_1}y1n}, together with $P_{x_2}=1-P_{x_1}$ and $P_{y_2}=1-P_{y_1}$, it is possible to solve \eqref{eq:Gamma}, finding the general expression of $\Gamma$.~
\begin{align} \label{eq:dergamma}   
\Gamma&=\frac{P_{x_1y_1} - P_{x_1}P_{y_1}}{P_{x_1}(1 - P_{x_1})},
\end{align}
which results in a function independent of the growth rates.~The mutual information $I$ can be obtained in a straightforward way by replacing the distributions explained in \eqref{eq:P_{x_1}y1n} in the formula \eqref{info}.~

Fig.~\ref{fig:generalized} shows two scenarios that can be distinguished depending on the marginal distributions: $P_{y_1}<P_{x_1}$ (Fig.~\ref{fig:generalized}A) and $P_{y_1}>P_{x_1}$ (Fig.~\ref{fig:generalized}B).~In both cases, keeping $P_{x_1}$ and $P_{y_1}$ fixed, $\Gamma$ and $I$ increase with $P_{x_1y_1}$ as a parameter.~In the case $P_{y_1}<P_{x_1}$, marginal distributions with higher $P_{y_1}$ allow both higher $\Gamma$ and $I$ while in the case $P_{y_1}>P_{x_1}$ distributions with higher $P_{y_1}$ will decrease $\Gamma$ and $I$.~The black line in Fig.~\ref{fig:generalized}A,B shows the highest achievable $\Gamma$ and $I$ when $P_{x_1y_1}= \min(P_{x_1},P_{y_1})$.~

To better understand how these maximum achievable $\Gamma$ and $I$ depend on the marginal phenotype distribution $P_{y_1}$, for a particular $P_{x_1}$, we present Fig.~\ref{fig:generalized}C, D.~In the interval $P_{y_1}\leq P_{x_1}$ the joint distribution that maximizes $I$ and $\Gamma$ follows $P_{x_1y_1}=P_{x_1}$ and for $P_{y_1}\geq P_{x_1}$, the joint distribution is given by $P_{x_1y_1}=P_{y_1}$.~By evaluating $\Gamma$ over these intervals, we obtain the following expression:
\begin{equation}
  \begin{split}
    \Gamma_{max} &=
    \begin{cases}
      \frac{1 - P_{y_1}}{1 - P_{x_1}} & \text{if } P_{x_1}\leq P_{y_1} \\
      \\
      \frac{P_{y_1}}{P_{x_1}} & \text{if } P_{y_1}< P_{x_1},
    \end{cases}
  \end{split}
\end{equation}
which shows us that $\Gamma_{max}$ is a global maximum when $P_{y_1}=P_{x_1}$ (Fig.~\ref{fig:generalized}C).

 Following similar logic, the maximum mutual information for particular values of $P_{x_1}$ and $P_{y_1}$ results in the following function depending upon these two intervals,
\begin{align} I_{max}=
 \begin{cases}
(1 - P_{y_1}) \log_2\left(\frac{1}{1 - P_{x_1}} \right)+ P_{x_1} \log_2\left(\frac{1}{P_{y_1}} \right) \\
     + (P_{y_1}-P_{x_1})\log_2\left(\frac{P_{x_1} - P_{y_1}}{(P_{x_1}-1 ) P_{y_1}}\right), ~~~ \text{if } P_{x_1}\leq P_{y_1} \\
\\
P_{y_1} \log_2\left(\frac{1}{P_{x_1}} \right)+ (1-P_{x_1}) \log_2\left(\frac{1}{1-P_{y_1}} \right) \\
\end{cases}
\end{align}
By solving this numerically, it also gives us the maximum possible information when $P_{y_1}=P_{x_1}$, shown in Fig.~\ref{fig:generalized}D.~This maximum achievable mutual information corresponds to the entropy of the environment.

In summary, we prove how for given marginal distributions $P_{x_1}$ and $P_{y_1}$ of environment and phenotype, respectively, any increase in population fitness through an arbitrary strategy corresponds to a simultaneous increase in information, and vice versa.~The information and relative fitness are maximized when $P_{y_1}=P_{x_1}$.~An important remark is that, in the general case, the relationship between $\Gamma$ and $I$ does not collapse in a single curve, as shown in Fig.~\ref{fig:information} for the symmetric-fitness model.~Instead, depending on the particular values of $P_{x_1}$, $P_{y_1}$ and $P_{x_1y_1}$; $\Gamma$ and $I$ will have values within the region depicted in Fig.~\ref{fig:generalized} A,B.~This implies that $I$ and $\Gamma$ will not necessarily increase together when $P_{x_1}$ and $P_{y_1}$ are not fixed.~We will explore this case after a short discussion about how information and fitness are related.

\subsection{The marginal information value}

In the previous section, we concluded that a key property of the relationship between mutual information $I$ and normalized population fitness $\Gamma$ is a monotonic increase:~with fixed marginal distributions, higher information consistently correlates with higher fitness.~This implies that an increase in information will also lead to an increase in normalized population fitness.~However, a small increase in information does not guarantee a proportional rise in population fitness.~To understand this marginal effect, we examine the partial derivative of the normalized growth rate with respect to mutual information keeping the marginal probabilities fixed.~This value, which we denote as the \textit{marginal information value}, quantifies the gain in normalized fitness achieved by a system with information $I$ when its information is increased infinitesimally.~From the expression \eqref{eq:dergamma}, we obtain the formula in terms of $P_{x_1}$, $P_{y_1}$ and $P_{x_1y_1}$:
\begin{align}\label{eq:value}
    \left(\frac{\partial \Gamma}{\partial I}\right)_{P_{x_1},P_{y_1}} = \frac{1}{(-1 + P_{x_1}) P_{x_1}} \left[
    \log_2\left(\frac{P_{x_1y_1} - P_{x_1}}{P_{x_1} (-1 + P_{y_1})}\right) \right.~\nonumber\\
    \left.~- \log_2\left(\frac{1 + P_{x_1y_1} - P_{x_1} - P_{y_1}}{(-1 + P_{x_1})(-1 + P_{y_1})}\right) \right.~\nonumber\\
    \left.~- \log_2\left(\frac{P_{x_1y_1}}{P_{x_1} P_{y_1}}\right) + \log_2\left(\frac{P_{x_1y_1} - P_{y_1}}{(-1 + P_{x_1}) P_{y_1}}\right) \right]^{-1}.
\end{align}

\begin{figure}[ht!]
    
\includegraphics[width=0.9\linewidth] {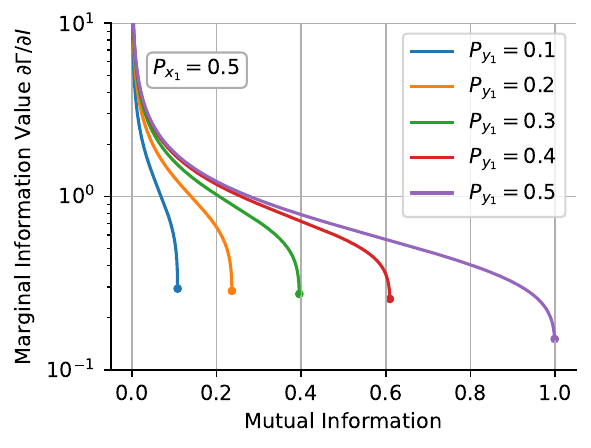}
    \caption{\textbf{The value of additional information decreases with the total mutual information for fixed $P_x$ and $P_y$.} The partial derivative of the normalized growth rate respective to the Information keeping constant $P_x$ and $P_y$ is interpreted as the marginal information value.~It is a decreasing function of the total information meaning that as the system gains information, the value of new information decreases.~For this plot, we consider an environment with distribution $P_{x_1}=0.5$.
    }\label{fig:Marginal}
\end{figure}

\begin{figure*}[ht!]
   \includegraphics[width=\linewidth]{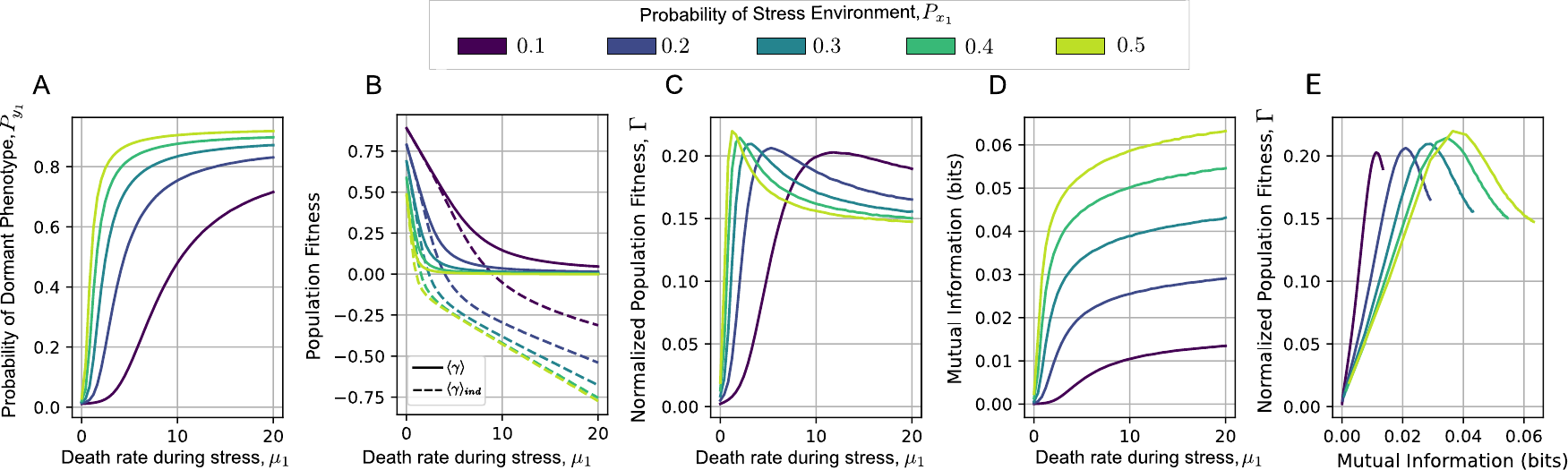}
    \caption{\textbf{Information and fitness in the model of dormant-proliferating phenotypes.} 
    Phenotype distribution $P_y$ can change as response of an environment with probability $P_x$ by increase parameters such as the relative growth rate.~For system of dormant-proliferating phenotypes, we increase the antibiotic killing rate $\mu_1$ relative to the proliferating rate $\mu_2$.
    \textbf{A.} By increasing $\mu_1$ for particular environmental distributions $P_{x_1}\in\{0.1,0.2,0.3,0.4,0.5\}$, the proportion of dormant cells $P_{y_1}$ will increase.~\textbf{B.} At the same time, the population fitness $\langle\gamma\rangle$ will decrease (solid line) but not as strongly as the population fitness if the phenotype is independent of the environmental state $\langle\gamma\rangle_{ind}$ (dashed line).~\textbf{C.} As a result, the normalized fitness, proportional to the difference between these population fitness will increase as $\mu_1$ increases reaching a maximal value when $P_{y_1}=P_{x_1}$ and decreasing as $P_{y_1}$ further increases.~\textbf{D.} The mutual information increases monotonically with $\mu_1$ meaning that the population needs more information to proliferate in a more selective environment.~\textbf{E.} Trend of normalized population fitness vs mutual information for different $P_{x_1}$ increasing $\mu_1$.~(Parameters: $\mu_2=1$, $k=0.01$, $\lambda_{21}=0.5$, $\lambda_{12}=\lambda_{21}\frac{1-P_{x_1}}{P_{x_1}}$).~ Time averages are estimated for a single trajectory of $f_{y_1}$, solution of \eqref{fraction_y1}, integrated over the time-span needed for performing 10 million of environmental transitions with initial condition $f_{y_1}|_{t=0}=0.5$.~}
    \label{fig:persistence}
\end{figure*}

In Fig.~\ref{fig:Marginal}, we show how the information value as given by \eqref{eq:value}, decreases as mutual information increases.~Initially, when the system has no information, the marginal information value is infinite.~This is because even an infinitesimal increase in information significantly boosts population fitness.~However, as the total information increases, each incremental gain in information contributes progressively less to the improvement of the population fitness.~Consequently, the value of newly acquired information decreases as the system has more knowledge.~Despite this decrease, the value never reaches zero, indicating that new information remains always valuable.~A more comprehensive approach would include the cost of increasing the system's information.~For instance, considering that increasing the responsive switching rate will result on a lower growth rate \cite{biswas2021gene}.~This interpretation will need the specific biologically context and would differ between systems \cite{lynch2015bioenergetic,mehta2016landauer}.

\subsection{Information and fitness: the model of dormant-proliferating phenotypes}

After discussing the general properties of information and normalized fitness, we explore how these variables can reveal different features of the population response in a general, complex, and relevant example.~As a final instance, we will study the model of persistent-proliferating phenotypes.~Among the multiple ways to analyze the system, we will observe how the explained variables change as the death rate during stress, $\mu_1$ in Table \ref{tab:persistence_growth_payoff} is increased while the other biological variables remain fixed:~the growth rate of the sensitive phenotype in proliferating conditions $\mu_2$, the phenotypical switching rate $k$, the rate of transition from stress to proliferating environment $\lambda_{21}$, and the probability of the proliferating environment $P_{x_1}$.~This approach can help us to understand, first, the importance of sharing information with the environment as is becomes more selective, and how much this information can improve the population success relative to the case of not sharing information.~As $P_{x_1}$ and $P_{y_1}$ are not fixed, we do not expect that the relationship between $I$ and $\Gamma$ follows a master curve as in the previous cases.

First, we observe the relationship between the marginal distribution of the dormant phenotypical state $P_{y_1}$, and the death rate during stress, $\mu_1$ (Fig.~\ref{fig:persistence}A).~As expected, $P_{y_1}$ increases with $\mu_1$, and this increase is more pronounced when the probability of encountering the stress environment, $P_{x_1}$, is higher, indicating stronger selective pressures.~Next, Fig.~\ref{fig:persistence}B compares the average population fitness obtained with bet-hedging, $\langle \gamma\rangle$ (solid line), to the fitness of a population with the same marginal distribution but independent variables, $\langle \gamma\rangle_{ind}$ (dashed line).~While $\langle \gamma\rangle$ decreases as $\mu_1$ increases, bet-hedging prevents it from becoming negative, thus avoiding population extinction.~In contrast, $\langle \gamma\rangle_{ind}$ can reach negative values at sufficiently high $\mu_1$.~The benefit conferred by bet-hedging is highlighted by the increasing gap between $\langle \gamma\rangle$ and $\langle \gamma\rangle_{ind}$ with increasing $\mu_1$.~This gap is quantified by the metric $\Gamma$ as shown in Fig.~\ref{fig:persistence}C.~Notably, $\Gamma$ is close to zero for low $\mu_1$ values, where the environment is less selective.~As $\mu_1$ increases, $\Gamma$ also increases, reaching a maximum value, and then decreases.~This maximum occurs when ${P_{y_1}\approx0.5}$ precisely when most individuals start to present the dormant phenotype.

Finally, we observe that, unlike $\Gamma$, mutual information increases monotonically with $\mu_1$.~This trend highlights the crucial role of increasing information in ensuring population survival under increasingly selective environmental pressures.~As a consequence of the differing responses of $\Gamma$ and $I$ as $\mu_1$ changes, their relationship is non-monotonic, with $\Gamma$ exhibiting a global maximum.~This non-monotonic relationship does not contradict our earlier conclusions, as $P_{y_1}$ does not remain fixed with increasing $\mu_1$.~This allowed $\Gamma$ and $I$ to exhibit any non-monotonic relationship within the regions depicted in Fig.~\ref{fig:generalized}A and B.~Taken together, these results offer different, yet complementary perspectives on how $\Gamma$ and $I$ can serve as descriptors of general population adaptation through both proliferation mechanisms and phenotypical plasticity.



\section{Discussion}

Information and fitness have been closely related in biosciences since the development of information theory~\cite{cohen1966optimizing}.~However, the mechanisms by which individuals encode and prioritize environmental information remains under debate~\cite{seoane2018information}.~Biological information processing is distinguished from classical communication channels by intrinsic variables such as self-replication, inherent noise in biological information pathways, and finite response times.~In biological systems, information flows through diverse biochemical pathways within organisms, including gene expression, signaling cascades, and neural systems~\cite{cheong2011information,hahn2023dynamical}.~Prior work has examined the fidelity of information transmission along specific pathways in relation to individual fitness~\cite{rivoire2011value,taylor2007information,donaldson2010fitness}, where an individual's phenotype, modeled as a random variable, represents an internal state related to the interpretation of the environment.~In contrast, our contribution investigates fitness as a population-level phenomenon.~From this viewpoint, competition favors the proliferation of the fittest phenotype, leading to its over-representation in the population.~Nevertheless, bet-hedging strategies and phenotypic noise prevent the fixation of this single fittest type, maintaining phenotypic diversity within the population.~This population-based perspective on fitness, considering phenotypic advantages, offers a complementary view to traditional single-individual information transmission.

It is important to acknowledge that the relevance of this population-based approach is context-dependent and may be superseded by individual fitness considerations in certain scenarios.~Our population-level interpretation gains prominence when proliferation rates are faster than the rates of environmental change.~The individual-based approach aligns with our population-based framework in the limit where environmental fluctuations are much faster than proliferation.~In such rapid environmental changes, individuals should have a timely and accurate representation of the environment.~Therefore, some individual-based approximations are justifiable.~These include modeling the environment and individual phenotypes as random variables rather than stochastic~processes and the information transmission as a process with no delay~\cite{moffett2022minimal,rivoire2011value,donaldson2010fitness}.


We define the phenotype and environmental probabilities as time averages of their associated stochastic processes (phenotype fraction and environmental state respectively)~\cite{skanata2016evolutionary}.~These probabilities are used to obtain the fitness of the population and the mutual information between the environment and the population phenotype.~We propose to measure the population fitness relative to the average growth rate when the population distribution is independent of the environment.~This normalized fitness does not dependent of the particular growth rate values and is an increasing function of information for fixed marginal distributions.~

In the symmetric bet-hedging model (Fig.~\ref{fig:information}) we observe how, regardless of the particular values of the growth rates, normalized fitness and mutual information collapse in a single curve that is a particular case for a symmetric environment: ${P_{x_1}=P_{y_1}=0.5}$.~It is possible to show that the environment with any general $P_{x_1}$ will generate other curves with similar properties.~Populations can achieve different points on that curve by having particular growth and phenotype switching rates.

We explore the properties of population and fitness in the general scenario with arbitrary switching and growth rates.~Without specifying these rates, we observe how fixing the values of the environmental state (input) $P_{x_1}$ and the phenotypical distribution (output) $P_{y_1}$, the normalized fitness is always an increasing function of mutual information.~We quantify the maximum achievable fitness and information for given values of $P_{x_1}$ and $P_{y_1}$ which occurs when $P_{x1y1}=\min(P_{x_1},P_{y_1})$.~In general, we observe that a strategy that maximizes information and fitness should satisfy: $P_{x_1}=P_{y_1}$.

A biologically relevant example is the model of dormant-proliferating phenotypes.~In this scenario, we observe two interesting properties; first, the output phenotype distribution depends on parameters such as the growth rates in each environment.~With an increase in the killing rate of antibiotics, the proportion of dormant phenotype increases for fixed environmental statistics.~When increasing this selection pressure, the information between the phenotypes and the environment increases, revealing the success of a bet-hedging strategy in this highly stressful context.~However, when increasing the pressure, the fitness of the total population decreases, but the fitness of the normalized population can increase and then decrease when the percentage of population with low growth rate becomes significant.~This behavior reveals how the normalized population fitness can exhibit properties between information and total population~fitness.

The relevance of acquiring information from a fluctuating environment in a heterogeneous population has been implicit in most of biological approaches \cite{donaldson2010fitness}.~In this article, we propose a more quantitative approach and observe that the purpose of acquiring information may be related to improving population performance relative to the hypothetical case of not acquiring information.~This relative performance, quantified by normalized fitness, can even exhibit trends different from the absolute population fitness.~Our proposed formalism provides more quantitative tools to understand the mechanisms that living beings use to adapt to environments.

\section*{Acknowledgments}
This article was funded by NIH-NIGMS through grant \texttt{R35GM148351}.

\section*{Data Accesibility Statement}
The codes used for our simulations can be found at https://doi.org/10.5281/zenodo.16427182.


\bibliographystyle{apsrev4-2}
\bibliography{references} 

\begin{thebibliography}{64}%
\makeatletter
\providecommand \@ifxundefined [1]{%
 \@ifx{#1\undefined}
}%
\providecommand \@ifnum [1]{%
 \ifnum #1\expandafter \@firstoftwo
 \else \expandafter \@secondoftwo
 \fi
}%
\providecommand \@ifx [1]{%
 \ifx #1\expandafter \@firstoftwo
 \else \expandafter \@secondoftwo
 \fi
}%
\providecommand \natexlab [1]{#1}%
\providecommand \enquote  [1]{``#1''}%
\providecommand \bibnamefont  [1]{#1}%
\providecommand \bibfnamefont [1]{#1}%
\providecommand \citenamefont [1]{#1}%
\providecommand \href@noop [0]{\@secondoftwo}%
\providecommand \href [0]{\begingroup \@sanitize@url \@href}%
\providecommand \@href[1]{\@@startlink{#1}\@@href}%
\providecommand \@@href[1]{\endgroup#1\@@endlink}%
\providecommand \@sanitize@url [0]{\catcode `\\12\catcode `\$12\catcode `\&12\catcode `\#12\catcode `\^12\catcode `\_12\catcode `\%12\relax}%
\providecommand \@@startlink[1]{}%
\providecommand \@@endlink[0]{}%
\providecommand \url  [0]{\begingroup\@sanitize@url \@url }%
\providecommand \@url [1]{\endgroup\@href {#1}{\urlprefix }}%
\providecommand \urlprefix  [0]{URL }%
\providecommand \Eprint [0]{\href }%
\providecommand \doibase [0]{https://doi.org/}%
\providecommand \selectlanguage [0]{\@gobble}%
\providecommand \bibinfo  [0]{\@secondoftwo}%
\providecommand \bibfield  [0]{\@secondoftwo}%
\providecommand \translation [1]{[#1]}%
\providecommand \BibitemOpen [0]{}%
\providecommand \bibitemStop [0]{}%
\providecommand \bibitemNoStop [0]{.\EOS\space}%
\providecommand \EOS [0]{\spacefactor3000\relax}%
\providecommand \BibitemShut  [1]{\csname bibitem#1\endcsname}%
\let\auto@bib@innerbib\@empty
\bibitem [{\citenamefont {Dekel}\ and\ \citenamefont {Alon}(2005)}]{dekel2005optimality}%
  \BibitemOpen
  \bibfield  {author} {\bibinfo {author} {\bibfnamefont {E.}~\bibnamefont {Dekel}}\ and\ \bibinfo {author} {\bibfnamefont {U.}~\bibnamefont {Alon}},\ }\href@noop {} {\bibfield  {journal} {\bibinfo  {journal} {Nature}\ }\textbf {\bibinfo {volume} {436}},\ \bibinfo {pages} {588} (\bibinfo {year} {2005})}\BibitemShut {NoStop}%
\bibitem [{\citenamefont {Aub{\'e}}\ and\ \citenamefont {Landry}(2024)}]{aube2024genotype}%
  \BibitemOpen
  \bibfield  {author} {\bibinfo {author} {\bibfnamefont {S.}~\bibnamefont {Aub{\'e}}}\ and\ \bibinfo {author} {\bibfnamefont {C.~R.}\ \bibnamefont {Landry}},\ }\href@noop {} {\bibfield  {journal} {\bibinfo  {journal} {Nature Ecology \& Evolution}\ ,\ \bibinfo {pages} {1}} (\bibinfo {year} {2024})}\BibitemShut {NoStop}%
\bibitem [{\citenamefont {Xue}\ \emph {et~al.}(2019)\citenamefont {Xue}, \citenamefont {Sartori},\ and\ \citenamefont {Leibler}}]{xue2019environment}%
  \BibitemOpen
  \bibfield  {author} {\bibinfo {author} {\bibfnamefont {B.}~\bibnamefont {Xue}}, \bibinfo {author} {\bibfnamefont {P.}~\bibnamefont {Sartori}},\ and\ \bibinfo {author} {\bibfnamefont {S.}~\bibnamefont {Leibler}},\ }\href@noop {} {\bibfield  {journal} {\bibinfo  {journal} {Proceedings of the National Academy of Sciences}\ }\textbf {\bibinfo {volume} {116}},\ \bibinfo {pages} {13847} (\bibinfo {year} {2019})}\BibitemShut {NoStop}%
\bibitem [{\citenamefont {De~Martino}\ \emph {et~al.}(2019)\citenamefont {De~Martino}, \citenamefont {Gueudr{\'e}},\ and\ \citenamefont {Miotto}}]{de2019exploration}%
  \BibitemOpen
  \bibfield  {author} {\bibinfo {author} {\bibfnamefont {A.}~\bibnamefont {De~Martino}}, \bibinfo {author} {\bibfnamefont {T.}~\bibnamefont {Gueudr{\'e}}},\ and\ \bibinfo {author} {\bibfnamefont {M.}~\bibnamefont {Miotto}},\ }\href@noop {} {\bibfield  {journal} {\bibinfo  {journal} {Physical Review E}\ }\textbf {\bibinfo {volume} {99}},\ \bibinfo {pages} {012417} (\bibinfo {year} {2019})}\BibitemShut {NoStop}%
\bibitem [{\citenamefont {Cvijovi{\'c}}\ \emph {et~al.}(2015)\citenamefont {Cvijovi{\'c}}, \citenamefont {Good}, \citenamefont {Jerison},\ and\ \citenamefont {Desai}}]{cvijovic2015fate}%
  \BibitemOpen
  \bibfield  {author} {\bibinfo {author} {\bibfnamefont {I.}~\bibnamefont {Cvijovi{\'c}}}, \bibinfo {author} {\bibfnamefont {B.~H.}\ \bibnamefont {Good}}, \bibinfo {author} {\bibfnamefont {E.~R.}\ \bibnamefont {Jerison}},\ and\ \bibinfo {author} {\bibfnamefont {M.~M.}\ \bibnamefont {Desai}},\ }\href@noop {} {\bibfield  {journal} {\bibinfo  {journal} {Proceedings of the National Academy of Sciences}\ }\textbf {\bibinfo {volume} {112}},\ \bibinfo {pages} {E5021} (\bibinfo {year} {2015})}\BibitemShut {NoStop}%
\bibitem [{\citenamefont {Levien}\ \emph {et~al.}(2021)\citenamefont {Levien}, \citenamefont {Min}, \citenamefont {Kondev},\ and\ \citenamefont {Amir}}]{levien2021non}%
  \BibitemOpen
  \bibfield  {author} {\bibinfo {author} {\bibfnamefont {E.}~\bibnamefont {Levien}}, \bibinfo {author} {\bibfnamefont {J.}~\bibnamefont {Min}}, \bibinfo {author} {\bibfnamefont {J.}~\bibnamefont {Kondev}},\ and\ \bibinfo {author} {\bibfnamefont {A.}~\bibnamefont {Amir}},\ }\href@noop {} {\bibfield  {journal} {\bibinfo  {journal} {Reports on Progress in Physics}\ }\textbf {\bibinfo {volume} {84}},\ \bibinfo {pages} {116601} (\bibinfo {year} {2021})}\BibitemShut {NoStop}%
\bibitem [{\citenamefont {de~Groot}\ \emph {et~al.}(2023)\citenamefont {de~Groot}, \citenamefont {Tjalma}, \citenamefont {Bruggeman},\ and\ \citenamefont {van Nimwegen}}]{de2023effective}%
  \BibitemOpen
  \bibfield  {author} {\bibinfo {author} {\bibfnamefont {D.~H.}\ \bibnamefont {de~Groot}}, \bibinfo {author} {\bibfnamefont {A.~J.}\ \bibnamefont {Tjalma}}, \bibinfo {author} {\bibfnamefont {F.~J.}\ \bibnamefont {Bruggeman}},\ and\ \bibinfo {author} {\bibfnamefont {E.}~\bibnamefont {van Nimwegen}},\ }\href@noop {} {\bibfield  {journal} {\bibinfo  {journal} {Proceedings of the National Academy of Sciences}\ }\textbf {\bibinfo {volume} {120}},\ \bibinfo {pages} {e2211091120} (\bibinfo {year} {2023})}\BibitemShut {NoStop}%
\bibitem [{\citenamefont {Kussell}\ and\ \citenamefont {Leibler}(2005)}]{kussell2005phenotypic}%
  \BibitemOpen
  \bibfield  {author} {\bibinfo {author} {\bibfnamefont {E.}~\bibnamefont {Kussell}}\ and\ \bibinfo {author} {\bibfnamefont {S.}~\bibnamefont {Leibler}},\ }\href@noop {} {\bibfield  {journal} {\bibinfo  {journal} {Science}\ }\textbf {\bibinfo {volume} {309}},\ \bibinfo {pages} {2075} (\bibinfo {year} {2005})}\BibitemShut {NoStop}%
\bibitem [{\citenamefont {Thattai}\ and\ \citenamefont {Van~Oudenaarden}(2004)}]{thattai2004stochastic}%
  \BibitemOpen
  \bibfield  {author} {\bibinfo {author} {\bibfnamefont {M.}~\bibnamefont {Thattai}}\ and\ \bibinfo {author} {\bibfnamefont {A.}~\bibnamefont {Van~Oudenaarden}},\ }\href@noop {} {\bibfield  {journal} {\bibinfo  {journal} {Genetics}\ }\textbf {\bibinfo {volume} {167}},\ \bibinfo {pages} {523} (\bibinfo {year} {2004})}\BibitemShut {NoStop}%
\bibitem [{\citenamefont {Fern{\'a}ndez-Fern{\'a}ndez}\ \emph {et~al.}(2024)\citenamefont {Fern{\'a}ndez-Fern{\'a}ndez}, \citenamefont {Olivenza}, \citenamefont {Weyer}, \citenamefont {Singh}, \citenamefont {Casades{\'u}s},\ and\ \citenamefont {Antonia S{\'a}nchez-Romero}}]{fernandez2024evolution}%
  \BibitemOpen
  \bibfield  {author} {\bibinfo {author} {\bibfnamefont {R.}~\bibnamefont {Fern{\'a}ndez-Fern{\'a}ndez}}, \bibinfo {author} {\bibfnamefont {D.~R.}\ \bibnamefont {Olivenza}}, \bibinfo {author} {\bibfnamefont {E.}~\bibnamefont {Weyer}}, \bibinfo {author} {\bibfnamefont {A.}~\bibnamefont {Singh}}, \bibinfo {author} {\bibfnamefont {J.}~\bibnamefont {Casades{\'u}s}},\ and\ \bibinfo {author} {\bibfnamefont {M.}~\bibnamefont {Antonia S{\'a}nchez-Romero}},\ }\href@noop {} {\bibfield  {journal} {\bibinfo  {journal} {Proceedings of the National Academy of Sciences}\ }\textbf {\bibinfo {volume} {121}},\ \bibinfo {pages} {e2322371121} (\bibinfo {year} {2024})}\BibitemShut {NoStop}%
\bibitem [{\citenamefont {Morawska}\ \emph {et~al.}(2022)\citenamefont {Morawska}, \citenamefont {Hernandez-Valdes},\ and\ \citenamefont {Kuipers}}]{morawska2022diversity}%
  \BibitemOpen
  \bibfield  {author} {\bibinfo {author} {\bibfnamefont {L.~P.}\ \bibnamefont {Morawska}}, \bibinfo {author} {\bibfnamefont {J.~A.}\ \bibnamefont {Hernandez-Valdes}},\ and\ \bibinfo {author} {\bibfnamefont {O.~P.}\ \bibnamefont {Kuipers}},\ }\href@noop {} {\bibfield  {journal} {\bibinfo  {journal} {WIREs Mechanisms of Disease}\ }\textbf {\bibinfo {volume} {14}},\ \bibinfo {pages} {e1544} (\bibinfo {year} {2022})}\BibitemShut {NoStop}%
\bibitem [{\citenamefont {Hawkins}\ and\ \citenamefont {Smolke}(2006)}]{hawkins2006regulatory}%
  \BibitemOpen
  \bibfield  {author} {\bibinfo {author} {\bibfnamefont {K.~M.}\ \bibnamefont {Hawkins}}\ and\ \bibinfo {author} {\bibfnamefont {C.~D.}\ \bibnamefont {Smolke}},\ }\href@noop {} {\bibfield  {journal} {\bibinfo  {journal} {Journal of Biological Chemistry}\ }\textbf {\bibinfo {volume} {281}},\ \bibinfo {pages} {13485} (\bibinfo {year} {2006})}\BibitemShut {NoStop}%
\bibitem [{\citenamefont {Solopova}\ \emph {et~al.}(2014)\citenamefont {Solopova}, \citenamefont {Van~Gestel}, \citenamefont {Weissing}, \citenamefont {Bachmann}, \citenamefont {Teusink}, \citenamefont {Kok},\ and\ \citenamefont {Kuipers}}]{solopova2014bet}%
  \BibitemOpen
  \bibfield  {author} {\bibinfo {author} {\bibfnamefont {A.}~\bibnamefont {Solopova}}, \bibinfo {author} {\bibfnamefont {J.}~\bibnamefont {Van~Gestel}}, \bibinfo {author} {\bibfnamefont {F.~J.}\ \bibnamefont {Weissing}}, \bibinfo {author} {\bibfnamefont {H.}~\bibnamefont {Bachmann}}, \bibinfo {author} {\bibfnamefont {B.}~\bibnamefont {Teusink}}, \bibinfo {author} {\bibfnamefont {J.}~\bibnamefont {Kok}},\ and\ \bibinfo {author} {\bibfnamefont {O.~P.}\ \bibnamefont {Kuipers}},\ }\href@noop {} {\bibfield  {journal} {\bibinfo  {journal} {Proceedings of the National Academy of Sciences}\ }\textbf {\bibinfo {volume} {111}},\ \bibinfo {pages} {7427} (\bibinfo {year} {2014})}\BibitemShut {NoStop}%
\bibitem [{\citenamefont {Maslov}\ and\ \citenamefont {Sneppen}(2015)}]{maslov2015well}%
  \BibitemOpen
  \bibfield  {author} {\bibinfo {author} {\bibfnamefont {S.}~\bibnamefont {Maslov}}\ and\ \bibinfo {author} {\bibfnamefont {K.}~\bibnamefont {Sneppen}},\ }\href@noop {} {\bibfield  {journal} {\bibinfo  {journal} {Scientific Reports}\ }\textbf {\bibinfo {volume} {5}},\ \bibinfo {pages} {10523} (\bibinfo {year} {2015})}\BibitemShut {NoStop}%
\bibitem [{\citenamefont {Bonifield}\ and\ \citenamefont {Hughes}(2003)}]{bonifield2003flagellar}%
  \BibitemOpen
  \bibfield  {author} {\bibinfo {author} {\bibfnamefont {H.~R.}\ \bibnamefont {Bonifield}}\ and\ \bibinfo {author} {\bibfnamefont {K.~T.}\ \bibnamefont {Hughes}},\ }\href@noop {} {\bibfield  {journal} {\bibinfo  {journal} {Journal of Bacteriology}\ }\textbf {\bibinfo {volume} {185}},\ \bibinfo {pages} {3567} (\bibinfo {year} {2003})}\BibitemShut {NoStop}%
\bibitem [{\citenamefont {Kenkel}\ and\ \citenamefont {Matz}(2016)}]{kenkel2016gene}%
  \BibitemOpen
  \bibfield  {author} {\bibinfo {author} {\bibfnamefont {C.~D.}\ \bibnamefont {Kenkel}}\ and\ \bibinfo {author} {\bibfnamefont {M.~V.}\ \bibnamefont {Matz}},\ }\href@noop {} {\bibfield  {journal} {\bibinfo  {journal} {Nature Ecology \& Evolution}\ }\textbf {\bibinfo {volume} {1}},\ \bibinfo {pages} {0014} (\bibinfo {year} {2016})}\BibitemShut {NoStop}%
\bibitem [{\citenamefont {Grissom}\ \emph {et~al.}(2024)\citenamefont {Grissom}, \citenamefont {Dixon}, \citenamefont {Singh},\ and\ \citenamefont {Blenner}}]{grissom2024heritable}%
  \BibitemOpen
  \bibfield  {author} {\bibinfo {author} {\bibfnamefont {S.}~\bibnamefont {Grissom}}, \bibinfo {author} {\bibfnamefont {Z.}~\bibnamefont {Dixon}}, \bibinfo {author} {\bibfnamefont {A.}~\bibnamefont {Singh}},\ and\ \bibinfo {author} {\bibfnamefont {M.}~\bibnamefont {Blenner}},\ }\href@noop {} {\bibfield  {journal} {\bibinfo  {journal} {bioRxiv}\ ,\ \bibinfo {pages} {2024}} (\bibinfo {year} {2024})}\BibitemShut {NoStop}%
\bibitem [{\citenamefont {Balaban}\ \emph {et~al.}(2004)\citenamefont {Balaban}, \citenamefont {Merrin}, \citenamefont {Chait}, \citenamefont {Kowalik},\ and\ \citenamefont {Leibler}}]{balaban2004bacterial}%
  \BibitemOpen
  \bibfield  {author} {\bibinfo {author} {\bibfnamefont {N.~Q.}\ \bibnamefont {Balaban}}, \bibinfo {author} {\bibfnamefont {J.}~\bibnamefont {Merrin}}, \bibinfo {author} {\bibfnamefont {R.}~\bibnamefont {Chait}}, \bibinfo {author} {\bibfnamefont {L.}~\bibnamefont {Kowalik}},\ and\ \bibinfo {author} {\bibfnamefont {S.}~\bibnamefont {Leibler}},\ }\href@noop {} {\bibfield  {journal} {\bibinfo  {journal} {Science}\ }\textbf {\bibinfo {volume} {305}},\ \bibinfo {pages} {1622} (\bibinfo {year} {2004})}\BibitemShut {NoStop}%
\bibitem [{\citenamefont {Patra}\ and\ \citenamefont {Klumpp}(2013)}]{patra2013population}%
  \BibitemOpen
  \bibfield  {author} {\bibinfo {author} {\bibfnamefont {P.}~\bibnamefont {Patra}}\ and\ \bibinfo {author} {\bibfnamefont {S.}~\bibnamefont {Klumpp}},\ }\href@noop {} {\bibfield  {journal} {\bibinfo  {journal} {PLoS One}\ }\textbf {\bibinfo {volume} {8}},\ \bibinfo {pages} {e62814} (\bibinfo {year} {2013})}\BibitemShut {NoStop}%
\bibitem [{\citenamefont {Hernandez-Beltran}\ \emph {et~al.}(2024)\citenamefont {Hernandez-Beltran}, \citenamefont {Rodr{\'\i}guez-Beltr{\'a}n}, \citenamefont {Aguilar-Luviano}, \citenamefont {Velez-Santiago}, \citenamefont {Mondrag{\'o}n-Palomino}, \citenamefont {MacLean}, \citenamefont {Fuentes-Hern{\'a}ndez}, \citenamefont {San~Mill{\'a}n},\ and\ \citenamefont {Pe{\~n}a-Miller}}]{hernandez2024plasmid}%
  \BibitemOpen
  \bibfield  {author} {\bibinfo {author} {\bibfnamefont {J.~C.~R.}\ \bibnamefont {Hernandez-Beltran}}, \bibinfo {author} {\bibfnamefont {J.}~\bibnamefont {Rodr{\'\i}guez-Beltr{\'a}n}}, \bibinfo {author} {\bibfnamefont {O.~B.}\ \bibnamefont {Aguilar-Luviano}}, \bibinfo {author} {\bibfnamefont {J.}~\bibnamefont {Velez-Santiago}}, \bibinfo {author} {\bibfnamefont {O.}~\bibnamefont {Mondrag{\'o}n-Palomino}}, \bibinfo {author} {\bibfnamefont {R.~C.}\ \bibnamefont {MacLean}}, \bibinfo {author} {\bibfnamefont {A.}~\bibnamefont {Fuentes-Hern{\'a}ndez}}, \bibinfo {author} {\bibfnamefont {A.}~\bibnamefont {San~Mill{\'a}n}},\ and\ \bibinfo {author} {\bibfnamefont {R.}~\bibnamefont {Pe{\~n}a-Miller}},\ }\href@noop {} {\bibfield  {journal} {\bibinfo  {journal} {Nature Communications}\ }\textbf {\bibinfo {volume} {15}},\ \bibinfo {pages} {2610} (\bibinfo {year} {2024})}\BibitemShut {NoStop}%
\bibitem [{\citenamefont {Diamond}\ and\ \citenamefont {Martin}(2021)}]{diamond2021buying}%
  \BibitemOpen
  \bibfield  {author} {\bibinfo {author} {\bibfnamefont {S.~E.}\ \bibnamefont {Diamond}}\ and\ \bibinfo {author} {\bibfnamefont {R.~A.}\ \bibnamefont {Martin}},\ }in\ \href@noop {} {\emph {\bibinfo {booktitle} {Phenotypic plasticity \& evolution}}}\ (\bibinfo  {publisher} {CRC Press},\ \bibinfo {year} {2021})\ pp.\ \bibinfo {pages} {185--209}\BibitemShut {NoStop}%
\bibitem [{\citenamefont {Rahman}\ \emph {et~al.}(2025)\citenamefont {Rahman}, \citenamefont {Amaratunga}, \citenamefont {Butzin}, \citenamefont {Singh}, \citenamefont {Hossain},\ and\ \citenamefont {Butzin}}]{rahman2025rethinking}%
  \BibitemOpen
  \bibfield  {author} {\bibinfo {author} {\bibfnamefont {K.~T.}\ \bibnamefont {Rahman}}, \bibinfo {author} {\bibfnamefont {R.}~\bibnamefont {Amaratunga}}, \bibinfo {author} {\bibfnamefont {X.~Y.}\ \bibnamefont {Butzin}}, \bibinfo {author} {\bibfnamefont {A.}~\bibnamefont {Singh}}, \bibinfo {author} {\bibfnamefont {T.}~\bibnamefont {Hossain}},\ and\ \bibinfo {author} {\bibfnamefont {N.~C.}\ \bibnamefont {Butzin}},\ }\href@noop {} {\bibfield  {journal} {\bibinfo  {journal} {International Journal of Antimicrobial Agents}\ }\textbf {\bibinfo {volume} {65}},\ \bibinfo {pages} {107386} (\bibinfo {year} {2025})}\BibitemShut {NoStop}%
\bibitem [{\citenamefont {Shaffer}\ \emph {et~al.}(2017)\citenamefont {Shaffer}, \citenamefont {Dunagin}, \citenamefont {Torborg}, \citenamefont {Torre}, \citenamefont {Emert}, \citenamefont {Krepler}, \citenamefont {Beqiri}, \citenamefont {Sproesser}, \citenamefont {Brafford}, \citenamefont {Xiao} \emph {et~al.}}]{shaffer2017rare}%
  \BibitemOpen
  \bibfield  {author} {\bibinfo {author} {\bibfnamefont {S.~M.}\ \bibnamefont {Shaffer}}, \bibinfo {author} {\bibfnamefont {M.~C.}\ \bibnamefont {Dunagin}}, \bibinfo {author} {\bibfnamefont {S.~R.}\ \bibnamefont {Torborg}}, \bibinfo {author} {\bibfnamefont {E.~A.}\ \bibnamefont {Torre}}, \bibinfo {author} {\bibfnamefont {B.}~\bibnamefont {Emert}}, \bibinfo {author} {\bibfnamefont {C.}~\bibnamefont {Krepler}}, \bibinfo {author} {\bibfnamefont {M.}~\bibnamefont {Beqiri}}, \bibinfo {author} {\bibfnamefont {K.}~\bibnamefont {Sproesser}}, \bibinfo {author} {\bibfnamefont {P.~A.}\ \bibnamefont {Brafford}}, \bibinfo {author} {\bibfnamefont {M.}~\bibnamefont {Xiao}}, \emph {et~al.},\ }\href@noop {} {\bibfield  {journal} {\bibinfo  {journal} {Nature}\ }\textbf {\bibinfo {volume} {546}},\ \bibinfo {pages} {431} (\bibinfo {year} {2017})}\BibitemShut {NoStop}%
\bibitem [{\citenamefont {Chang}\ \emph {et~al.}(2022)\citenamefont {Chang}, \citenamefont {Jen}, \citenamefont {Jiang}, \citenamefont {Sayad}, \citenamefont {Mer}, \citenamefont {Brown}, \citenamefont {Nixon}, \citenamefont {Dhabaria}, \citenamefont {Tang}, \citenamefont {Venet} \emph {et~al.}}]{chang2022ontogeny}%
  \BibitemOpen
  \bibfield  {author} {\bibinfo {author} {\bibfnamefont {C.~A.}\ \bibnamefont {Chang}}, \bibinfo {author} {\bibfnamefont {J.}~\bibnamefont {Jen}}, \bibinfo {author} {\bibfnamefont {S.}~\bibnamefont {Jiang}}, \bibinfo {author} {\bibfnamefont {A.}~\bibnamefont {Sayad}}, \bibinfo {author} {\bibfnamefont {A.~S.}\ \bibnamefont {Mer}}, \bibinfo {author} {\bibfnamefont {K.~R.}\ \bibnamefont {Brown}}, \bibinfo {author} {\bibfnamefont {A.~M.}\ \bibnamefont {Nixon}}, \bibinfo {author} {\bibfnamefont {A.}~\bibnamefont {Dhabaria}}, \bibinfo {author} {\bibfnamefont {K.~H.}\ \bibnamefont {Tang}}, \bibinfo {author} {\bibfnamefont {D.}~\bibnamefont {Venet}}, \emph {et~al.},\ }\href@noop {} {\bibfield  {journal} {\bibinfo  {journal} {Cancer Discovery}\ }\textbf {\bibinfo {volume} {12}},\ \bibinfo {pages} {1022} (\bibinfo {year} {2022})}\BibitemShut {NoStop}%
\bibitem [{\citenamefont {Harmange}\ \emph {et~al.}(2023)\citenamefont {Harmange}, \citenamefont {Hueros}, \citenamefont {Schaff}, \citenamefont {Emert}, \citenamefont {Saint-Antoine}, \citenamefont {Kim}, \citenamefont {Niu}, \citenamefont {Nellore}, \citenamefont {Fane}, \citenamefont {Alicea} \emph {et~al.}}]{harmange2023disrupting}%
  \BibitemOpen
  \bibfield  {author} {\bibinfo {author} {\bibfnamefont {G.}~\bibnamefont {Harmange}}, \bibinfo {author} {\bibfnamefont {R.~A.~R.}\ \bibnamefont {Hueros}}, \bibinfo {author} {\bibfnamefont {D.~L.}\ \bibnamefont {Schaff}}, \bibinfo {author} {\bibfnamefont {B.}~\bibnamefont {Emert}}, \bibinfo {author} {\bibfnamefont {M.}~\bibnamefont {Saint-Antoine}}, \bibinfo {author} {\bibfnamefont {L.~C.}\ \bibnamefont {Kim}}, \bibinfo {author} {\bibfnamefont {Z.}~\bibnamefont {Niu}}, \bibinfo {author} {\bibfnamefont {S.}~\bibnamefont {Nellore}}, \bibinfo {author} {\bibfnamefont {M.~E.}\ \bibnamefont {Fane}}, \bibinfo {author} {\bibfnamefont {G.~M.}\ \bibnamefont {Alicea}}, \emph {et~al.},\ }\href@noop {} {\bibfield  {journal} {\bibinfo  {journal} {Nature Communications}\ }\textbf {\bibinfo {volume} {14}},\ \bibinfo {pages} {7130} (\bibinfo {year} {2023})}\BibitemShut {NoStop}%
\bibitem [{\citenamefont {Henrion}\ \emph {et~al.}(2023)\citenamefont {Henrion}, \citenamefont {Martinez}, \citenamefont {Vandenbroucke}, \citenamefont {Delvenne}, \citenamefont {Telek}, \citenamefont {Zicler}, \citenamefont {Gr{\"u}nberger},\ and\ \citenamefont {Delvigne}}]{henrion2023fitness}%
  \BibitemOpen
  \bibfield  {author} {\bibinfo {author} {\bibfnamefont {L.}~\bibnamefont {Henrion}}, \bibinfo {author} {\bibfnamefont {J.~A.}\ \bibnamefont {Martinez}}, \bibinfo {author} {\bibfnamefont {V.}~\bibnamefont {Vandenbroucke}}, \bibinfo {author} {\bibfnamefont {M.}~\bibnamefont {Delvenne}}, \bibinfo {author} {\bibfnamefont {S.}~\bibnamefont {Telek}}, \bibinfo {author} {\bibfnamefont {A.}~\bibnamefont {Zicler}}, \bibinfo {author} {\bibfnamefont {A.}~\bibnamefont {Gr{\"u}nberger}},\ and\ \bibinfo {author} {\bibfnamefont {F.}~\bibnamefont {Delvigne}},\ }\href@noop {} {\bibfield  {journal} {\bibinfo  {journal} {Nature Communications}\ }\textbf {\bibinfo {volume} {14}},\ \bibinfo {pages} {6128} (\bibinfo {year} {2023})}\BibitemShut {NoStop}%
\bibitem [{\citenamefont {Josephs}(2018)}]{josephs2018determining}%
  \BibitemOpen
  \bibfield  {author} {\bibinfo {author} {\bibfnamefont {E.~B.}\ \bibnamefont {Josephs}},\ }\href@noop {} {\bibfield  {journal} {\bibinfo  {journal} {New Phytologist}\ }\textbf {\bibinfo {volume} {219}},\ \bibinfo {pages} {31} (\bibinfo {year} {2018})}\BibitemShut {NoStop}%
\bibitem [{\citenamefont {Vahdat}\ \emph {et~al.}(2024)\citenamefont {Vahdat}, \citenamefont {Rezaee},\ and\ \citenamefont {Singh}}]{vahdat2024capturing}%
  \BibitemOpen
  \bibfield  {author} {\bibinfo {author} {\bibfnamefont {Z.}~\bibnamefont {Vahdat}}, \bibinfo {author} {\bibfnamefont {S.}~\bibnamefont {Rezaee}},\ and\ \bibinfo {author} {\bibfnamefont {A.}~\bibnamefont {Singh}},\ }\href@noop {} {\bibfield  {journal} {\bibinfo  {journal} {IFAC-PapersOnLine}\ }\textbf {\bibinfo {volume} {58}},\ \bibinfo {pages} {276} (\bibinfo {year} {2024})}\BibitemShut {NoStop}%
\bibitem [{\citenamefont {Zhang}\ \emph {et~al.}(2025)\citenamefont {Zhang}, \citenamefont {Zabaikina}, \citenamefont {Nieto}, \citenamefont {Vahdat}, \citenamefont {Bokes},\ and\ \citenamefont {Singh}}]{zhang2025stochastic}%
  \BibitemOpen
  \bibfield  {author} {\bibinfo {author} {\bibfnamefont {Z.}~\bibnamefont {Zhang}}, \bibinfo {author} {\bibfnamefont {I.}~\bibnamefont {Zabaikina}}, \bibinfo {author} {\bibfnamefont {C.}~\bibnamefont {Nieto}}, \bibinfo {author} {\bibfnamefont {Z.}~\bibnamefont {Vahdat}}, \bibinfo {author} {\bibfnamefont {P.}~\bibnamefont {Bokes}},\ and\ \bibinfo {author} {\bibfnamefont {A.}~\bibnamefont {Singh}},\ }\href@noop {} {\bibfield  {journal} {\bibinfo  {journal} {PLOS Computational Biology}\ }\textbf {\bibinfo {volume} {21}},\ \bibinfo {pages} {e1013014} (\bibinfo {year} {2025})}\BibitemShut {NoStop}%
\bibitem [{\citenamefont {Siddiq}\ \emph {et~al.}(2024)\citenamefont {Siddiq}, \citenamefont {Duveau},\ and\ \citenamefont {Wittkopp}}]{siddiq2024plasticity}%
  \BibitemOpen
  \bibfield  {author} {\bibinfo {author} {\bibfnamefont {M.~A.}\ \bibnamefont {Siddiq}}, \bibinfo {author} {\bibfnamefont {F.}~\bibnamefont {Duveau}},\ and\ \bibinfo {author} {\bibfnamefont {P.~J.}\ \bibnamefont {Wittkopp}},\ }\href@noop {} {\bibfield  {journal} {\bibinfo  {journal} {Nature Ecology \& Evolution}\ ,\ \bibinfo {pages} {1}} (\bibinfo {year} {2024})}\BibitemShut {NoStop}%
\bibitem [{\citenamefont {Singh}\ and\ \citenamefont {Saint-Antoine}(2023)}]{singh2023probing}%
  \BibitemOpen
  \bibfield  {author} {\bibinfo {author} {\bibfnamefont {A.}~\bibnamefont {Singh}}\ and\ \bibinfo {author} {\bibfnamefont {M.}~\bibnamefont {Saint-Antoine}},\ }\href@noop {} {\bibfield  {journal} {\bibinfo  {journal} {Frontiers in Microbiology}\ }\textbf {\bibinfo {volume} {13}},\ \bibinfo {pages} {1050516} (\bibinfo {year} {2023})}\BibitemShut {NoStop}%
\bibitem [{\citenamefont {Mattingly}\ and\ \citenamefont {Emonet}(2022)}]{mattingly2022collective}%
  \BibitemOpen
  \bibfield  {author} {\bibinfo {author} {\bibfnamefont {H.~H.}\ \bibnamefont {Mattingly}}\ and\ \bibinfo {author} {\bibfnamefont {T.}~\bibnamefont {Emonet}},\ }\href@noop {} {\bibfield  {journal} {\bibinfo  {journal} {Proceedings of the National Academy of Sciences}\ }\textbf {\bibinfo {volume} {119}},\ \bibinfo {pages} {e2117377119} (\bibinfo {year} {2022})}\BibitemShut {NoStop}%
\bibitem [{\citenamefont {Scheiner}(1993)}]{scheiner1993genetics}%
  \BibitemOpen
  \bibfield  {author} {\bibinfo {author} {\bibfnamefont {S.~M.}\ \bibnamefont {Scheiner}},\ }\href@noop {} {\bibfield  {journal} {\bibinfo  {journal} {Annual Review of Ecology and Systematics}\ }\textbf {\bibinfo {volume} {24}},\ \bibinfo {pages} {35} (\bibinfo {year} {1993})}\BibitemShut {NoStop}%
\bibitem [{\citenamefont {Bernhardt}\ \emph {et~al.}(2020)\citenamefont {Bernhardt}, \citenamefont {O'Connor}, \citenamefont {Sunday},\ and\ \citenamefont {Gonzalez}}]{bernhardt2020life}%
  \BibitemOpen
  \bibfield  {author} {\bibinfo {author} {\bibfnamefont {J.~R.}\ \bibnamefont {Bernhardt}}, \bibinfo {author} {\bibfnamefont {M.~I.}\ \bibnamefont {O'Connor}}, \bibinfo {author} {\bibfnamefont {J.~M.}\ \bibnamefont {Sunday}},\ and\ \bibinfo {author} {\bibfnamefont {A.}~\bibnamefont {Gonzalez}},\ }\href@noop {} {\bibfield  {journal} {\bibinfo  {journal} {Philosophical Transactions of the Royal Society B}\ }\textbf {\bibinfo {volume} {375}},\ \bibinfo {pages} {20190454} (\bibinfo {year} {2020})}\BibitemShut {NoStop}%
\bibitem [{\citenamefont {Libby}\ \emph {et~al.}(2007)\citenamefont {Libby}, \citenamefont {Perkins},\ and\ \citenamefont {Swain}}]{libby2007noisy}%
  \BibitemOpen
  \bibfield  {author} {\bibinfo {author} {\bibfnamefont {E.}~\bibnamefont {Libby}}, \bibinfo {author} {\bibfnamefont {T.~J.}\ \bibnamefont {Perkins}},\ and\ \bibinfo {author} {\bibfnamefont {P.~S.}\ \bibnamefont {Swain}},\ }\href@noop {} {\bibfield  {journal} {\bibinfo  {journal} {Proceedings of the National Academy of Sciences}\ }\textbf {\bibinfo {volume} {104}},\ \bibinfo {pages} {7151} (\bibinfo {year} {2007})}\BibitemShut {NoStop}%
\bibitem [{\citenamefont {Koonin}(2016)}]{koonin2016meaning}%
  \BibitemOpen
  \bibfield  {author} {\bibinfo {author} {\bibfnamefont {E.~V.}\ \bibnamefont {Koonin}},\ }\href@noop {} {\bibfield  {journal} {\bibinfo  {journal} {Philosophical Transactions of the Royal Society A: Mathematical, Physical and Engineering Sciences}\ }\textbf {\bibinfo {volume} {374}},\ \bibinfo {pages} {20150065} (\bibinfo {year} {2016})}\BibitemShut {NoStop}%
\bibitem [{\citenamefont {Shannon}(2001)}]{shannon2001mathematical}%
  \BibitemOpen
  \bibfield  {author} {\bibinfo {author} {\bibfnamefont {C.~E.}\ \bibnamefont {Shannon}},\ }\href@noop {} {\bibfield  {journal} {\bibinfo  {journal} {Comput. Commun. Rev}\ }\textbf {\bibinfo {volume} {5}},\ \bibinfo {pages} {3} (\bibinfo {year} {2001})}\BibitemShut {NoStop}%
\bibitem [{\citenamefont {Bermek}(2024)}]{bermek2024form}%
  \BibitemOpen
  \bibfield  {author} {\bibinfo {author} {\bibfnamefont {E.}~\bibnamefont {Bermek}},\ }\href@noop {} {\bibfield  {journal} {\bibinfo  {journal} {Foundations of Science}\ ,\ \bibinfo {pages} {1}} (\bibinfo {year} {2024})}\BibitemShut {NoStop}%
\bibitem [{\citenamefont {Donaldson-Matasci}\ \emph {et~al.}(2010)\citenamefont {Donaldson-Matasci}, \citenamefont {Bergstrom},\ and\ \citenamefont {Lachmann}}]{donaldson2010fitness}%
  \BibitemOpen
  \bibfield  {author} {\bibinfo {author} {\bibfnamefont {M.~C.}\ \bibnamefont {Donaldson-Matasci}}, \bibinfo {author} {\bibfnamefont {C.~T.}\ \bibnamefont {Bergstrom}},\ and\ \bibinfo {author} {\bibfnamefont {M.}~\bibnamefont {Lachmann}},\ }\href@noop {} {\bibfield  {journal} {\bibinfo  {journal} {Oikos}\ }\textbf {\bibinfo {volume} {119}},\ \bibinfo {pages} {219} (\bibinfo {year} {2010})}\BibitemShut {NoStop}%
\bibitem [{\citenamefont {Hidalgo}\ \emph {et~al.}(2014)\citenamefont {Hidalgo}, \citenamefont {Grilli}, \citenamefont {Suweis}, \citenamefont {Munoz}, \citenamefont {Banavar},\ and\ \citenamefont {Maritan}}]{hidalgo2014information}%
  \BibitemOpen
  \bibfield  {author} {\bibinfo {author} {\bibfnamefont {J.}~\bibnamefont {Hidalgo}}, \bibinfo {author} {\bibfnamefont {J.}~\bibnamefont {Grilli}}, \bibinfo {author} {\bibfnamefont {S.}~\bibnamefont {Suweis}}, \bibinfo {author} {\bibfnamefont {M.~A.}\ \bibnamefont {Munoz}}, \bibinfo {author} {\bibfnamefont {J.~R.}\ \bibnamefont {Banavar}},\ and\ \bibinfo {author} {\bibfnamefont {A.}~\bibnamefont {Maritan}},\ }\href@noop {} {\bibfield  {journal} {\bibinfo  {journal} {Proceedings of the National Academy of Sciences}\ }\textbf {\bibinfo {volume} {111}},\ \bibinfo {pages} {10095} (\bibinfo {year} {2014})}\BibitemShut {NoStop}%
\bibitem [{\citenamefont {Pedraza}\ \emph {et~al.}(2018)\citenamefont {Pedraza}, \citenamefont {Garcia},\ and\ \citenamefont {P{\'e}rez-Ortiz}}]{pedraza2018noise}%
  \BibitemOpen
  \bibfield  {author} {\bibinfo {author} {\bibfnamefont {J.~M.}\ \bibnamefont {Pedraza}}, \bibinfo {author} {\bibfnamefont {D.~A.}\ \bibnamefont {Garcia}},\ and\ \bibinfo {author} {\bibfnamefont {M.~F.}\ \bibnamefont {P{\'e}rez-Ortiz}},\ }\href@noop {} {\bibfield  {journal} {\bibinfo  {journal} {Frontiers in Physics}\ }\textbf {\bibinfo {volume} {6}},\ \bibinfo {pages} {83} (\bibinfo {year} {2018})}\BibitemShut {NoStop}%
\bibitem [{\citenamefont {Moffett}\ and\ \citenamefont {Eckford}(2022)}]{moffett2022minimal}%
  \BibitemOpen
  \bibfield  {author} {\bibinfo {author} {\bibfnamefont {A.~S.}\ \bibnamefont {Moffett}}\ and\ \bibinfo {author} {\bibfnamefont {A.~W.}\ \bibnamefont {Eckford}},\ }\href@noop {} {\bibfield  {journal} {\bibinfo  {journal} {Physical Review E}\ }\textbf {\bibinfo {volume} {105}},\ \bibinfo {pages} {014403} (\bibinfo {year} {2022})}\BibitemShut {NoStop}%
\bibitem [{\citenamefont {Rivoire}\ and\ \citenamefont {Leibler}(2011)}]{rivoire2011value}%
  \BibitemOpen
  \bibfield  {author} {\bibinfo {author} {\bibfnamefont {O.}~\bibnamefont {Rivoire}}\ and\ \bibinfo {author} {\bibfnamefont {S.}~\bibnamefont {Leibler}},\ }\href@noop {} {\bibfield  {journal} {\bibinfo  {journal} {Journal of Statistical Physics}\ }\textbf {\bibinfo {volume} {142}},\ \bibinfo {pages} {1124} (\bibinfo {year} {2011})}\BibitemShut {NoStop}%
\bibitem [{\citenamefont {Taylor}\ \emph {et~al.}(2007)\citenamefont {Taylor}, \citenamefont {Tishby},\ and\ \citenamefont {Bialek}}]{taylor2007information}%
  \BibitemOpen
  \bibfield  {author} {\bibinfo {author} {\bibfnamefont {S.~F.}\ \bibnamefont {Taylor}}, \bibinfo {author} {\bibfnamefont {N.}~\bibnamefont {Tishby}},\ and\ \bibinfo {author} {\bibfnamefont {W.}~\bibnamefont {Bialek}},\ }\href@noop {} {\bibfield  {journal} {\bibinfo  {journal} {arXiv preprint arXiv:0712.4382}\ } (\bibinfo {year} {2007})}\BibitemShut {NoStop}%
\bibitem [{\citenamefont {Slein}\ \emph {et~al.}(2023)\citenamefont {Slein}, \citenamefont {Bernhardt}, \citenamefont {O'Connor},\ and\ \citenamefont {Fey}}]{slein2023effects}%
  \BibitemOpen
  \bibfield  {author} {\bibinfo {author} {\bibfnamefont {M.~A.}\ \bibnamefont {Slein}}, \bibinfo {author} {\bibfnamefont {J.~R.}\ \bibnamefont {Bernhardt}}, \bibinfo {author} {\bibfnamefont {M.~I.}\ \bibnamefont {O'Connor}},\ and\ \bibinfo {author} {\bibfnamefont {S.~B.}\ \bibnamefont {Fey}},\ }\href@noop {} {\bibfield  {journal} {\bibinfo  {journal} {Proceedings of the Royal Society B}\ }\textbf {\bibinfo {volume} {290}},\ \bibinfo {pages} {20222225} (\bibinfo {year} {2023})}\BibitemShut {NoStop}%
\bibitem [{\citenamefont {Vermeersch}\ \emph {et~al.}(2022)\citenamefont {Vermeersch}, \citenamefont {Cool}, \citenamefont {Gorkovskiy}, \citenamefont {Voordeckers}, \citenamefont {Wenseleers},\ and\ \citenamefont {Verstrepen}}]{vermeersch2022microbes}%
  \BibitemOpen
  \bibfield  {author} {\bibinfo {author} {\bibfnamefont {L.}~\bibnamefont {Vermeersch}}, \bibinfo {author} {\bibfnamefont {L.}~\bibnamefont {Cool}}, \bibinfo {author} {\bibfnamefont {A.}~\bibnamefont {Gorkovskiy}}, \bibinfo {author} {\bibfnamefont {K.}~\bibnamefont {Voordeckers}}, \bibinfo {author} {\bibfnamefont {T.}~\bibnamefont {Wenseleers}},\ and\ \bibinfo {author} {\bibfnamefont {K.~J.}\ \bibnamefont {Verstrepen}},\ }\href@noop {} {\bibfield  {journal} {\bibinfo  {journal} {Frontiers in Microbiology}\ }\textbf {\bibinfo {volume} {13}},\ \bibinfo {pages} {1004488} (\bibinfo {year} {2022})}\BibitemShut {NoStop}%
\bibitem [{\citenamefont {Skanata}\ and\ \citenamefont {Kussell}(2016)}]{skanata2016evolutionary}%
  \BibitemOpen
  \bibfield  {author} {\bibinfo {author} {\bibfnamefont {A.}~\bibnamefont {Skanata}}\ and\ \bibinfo {author} {\bibfnamefont {E.}~\bibnamefont {Kussell}},\ }\href@noop {} {\bibfield  {journal} {\bibinfo  {journal} {Physical review letters}\ }\textbf {\bibinfo {volume} {117}},\ \bibinfo {pages} {038104} (\bibinfo {year} {2016})}\BibitemShut {NoStop}%
\bibitem [{\citenamefont {D{\'a}vila-Romero}\ \emph {et~al.}(2024)\citenamefont {D{\'a}vila-Romero}, \citenamefont {Cao-Garc{\'\i}a},\ and\ \citenamefont {Dinis}}]{davila2024extinction}%
  \BibitemOpen
  \bibfield  {author} {\bibinfo {author} {\bibfnamefont {M.}~\bibnamefont {D{\'a}vila-Romero}}, \bibinfo {author} {\bibfnamefont {F.~J.}\ \bibnamefont {Cao-Garc{\'\i}a}},\ and\ \bibinfo {author} {\bibfnamefont {L.}~\bibnamefont {Dinis}},\ }\href@noop {} {\bibfield  {journal} {\bibinfo  {journal} {arXiv preprint arXiv:2406.11482}\ } (\bibinfo {year} {2024})}\BibitemShut {NoStop}%
\bibitem [{\citenamefont {Shin}\ and\ \citenamefont {Bleris}(2010)}]{shin2010linear}%
  \BibitemOpen
  \bibfield  {author} {\bibinfo {author} {\bibfnamefont {Y.-J.}\ \bibnamefont {Shin}}\ and\ \bibinfo {author} {\bibfnamefont {L.}~\bibnamefont {Bleris}},\ }\href@noop {} {\bibfield  {journal} {\bibinfo  {journal} {PloS one}\ }\textbf {\bibinfo {volume} {5}},\ \bibinfo {pages} {e12785} (\bibinfo {year} {2010})}\BibitemShut {NoStop}%
\bibitem [{\citenamefont {Van Der~Meer}\ and\ \citenamefont {Belkin}(2010)}]{van2010microbiology}%
  \BibitemOpen
  \bibfield  {author} {\bibinfo {author} {\bibfnamefont {J.~R.}\ \bibnamefont {Van Der~Meer}}\ and\ \bibinfo {author} {\bibfnamefont {S.}~\bibnamefont {Belkin}},\ }\href@noop {} {\bibfield  {journal} {\bibinfo  {journal} {Nature Reviews Microbiology}\ }\textbf {\bibinfo {volume} {8}},\ \bibinfo {pages} {511} (\bibinfo {year} {2010})}\BibitemShut {NoStop}%
\bibitem [{\citenamefont {Taheri-Araghi}\ \emph {et~al.}(2015)\citenamefont {Taheri-Araghi}, \citenamefont {Bradde}, \citenamefont {Sauls}, \citenamefont {Hill}, \citenamefont {Levin}, \citenamefont {Paulsson}, \citenamefont {Vergassola},\ and\ \citenamefont {Jun}}]{taheri2015cell}%
  \BibitemOpen
  \bibfield  {author} {\bibinfo {author} {\bibfnamefont {S.}~\bibnamefont {Taheri-Araghi}}, \bibinfo {author} {\bibfnamefont {S.}~\bibnamefont {Bradde}}, \bibinfo {author} {\bibfnamefont {J.~T.}\ \bibnamefont {Sauls}}, \bibinfo {author} {\bibfnamefont {N.~S.}\ \bibnamefont {Hill}}, \bibinfo {author} {\bibfnamefont {P.~A.}\ \bibnamefont {Levin}}, \bibinfo {author} {\bibfnamefont {J.}~\bibnamefont {Paulsson}}, \bibinfo {author} {\bibfnamefont {M.}~\bibnamefont {Vergassola}},\ and\ \bibinfo {author} {\bibfnamefont {S.}~\bibnamefont {Jun}},\ }\href@noop {} {\bibfield  {journal} {\bibinfo  {journal} {Current Biology}\ }\textbf {\bibinfo {volume} {25}},\ \bibinfo {pages} {385} (\bibinfo {year} {2015})}\BibitemShut {NoStop}%
\bibitem [{\citenamefont {Dur{\'a}n}\ \emph {et~al.}(2020)\citenamefont {Dur{\'a}n}, \citenamefont {Hern{\'a}ndez}, \citenamefont {Suesca}, \citenamefont {Acevedo}, \citenamefont {Acosta}, \citenamefont {Forero}, \citenamefont {Rozo},\ and\ \citenamefont {Pedraza}}]{duran2020slipstreaming}%
  \BibitemOpen
  \bibfield  {author} {\bibinfo {author} {\bibfnamefont {D.~C.}\ \bibnamefont {Dur{\'a}n}}, \bibinfo {author} {\bibfnamefont {C.~A.}\ \bibnamefont {Hern{\'a}ndez}}, \bibinfo {author} {\bibfnamefont {E.}~\bibnamefont {Suesca}}, \bibinfo {author} {\bibfnamefont {R.}~\bibnamefont {Acevedo}}, \bibinfo {author} {\bibfnamefont {I.~M.}\ \bibnamefont {Acosta}}, \bibinfo {author} {\bibfnamefont {D.~A.}\ \bibnamefont {Forero}}, \bibinfo {author} {\bibfnamefont {F.~E.}\ \bibnamefont {Rozo}},\ and\ \bibinfo {author} {\bibfnamefont {J.~M.}\ \bibnamefont {Pedraza}},\ }\href@noop {} {\bibfield  {journal} {\bibinfo  {journal} {Micromachines}\ }\textbf {\bibinfo {volume} {12}},\ \bibinfo {pages} {4} (\bibinfo {year} {2020})}\BibitemShut {NoStop}%
\bibitem [{\citenamefont {Guha}\ \emph {et~al.}(2025)\citenamefont {Guha}, \citenamefont {Singh},\ and\ \citenamefont {Butzin}}]{guha2025priestia}%
  \BibitemOpen
  \bibfield  {author} {\bibinfo {author} {\bibfnamefont {M.}~\bibnamefont {Guha}}, \bibinfo {author} {\bibfnamefont {A.}~\bibnamefont {Singh}},\ and\ \bibinfo {author} {\bibfnamefont {N.~C.}\ \bibnamefont {Butzin}},\ }\href@noop {} {\bibfield  {journal} {\bibinfo  {journal} {Communications Biology}\ }\textbf {\bibinfo {volume} {8}},\ \bibinfo {pages} {206} (\bibinfo {year} {2025})}\BibitemShut {NoStop}%
\bibitem [{\citenamefont {Hossain}\ \emph {et~al.}(2023)\citenamefont {Hossain}, \citenamefont {Singh},\ and\ \citenamefont {Butzin}}]{hossain2023escherichia}%
  \BibitemOpen
  \bibfield  {author} {\bibinfo {author} {\bibfnamefont {T.}~\bibnamefont {Hossain}}, \bibinfo {author} {\bibfnamefont {A.}~\bibnamefont {Singh}},\ and\ \bibinfo {author} {\bibfnamefont {N.~C.}\ \bibnamefont {Butzin}},\ }\href@noop {} {\bibfield  {journal} {\bibinfo  {journal} {Microbiology Spectrum}\ }\textbf {\bibinfo {volume} {11}},\ \bibinfo {pages} {e01219} (\bibinfo {year} {2023})}\BibitemShut {NoStop}%
\bibitem [{\citenamefont {Bokes}\ and\ \citenamefont {Singh}(2025)}]{bokes2025optimisation}%
  \BibitemOpen
  \bibfield  {author} {\bibinfo {author} {\bibfnamefont {P.}~\bibnamefont {Bokes}}\ and\ \bibinfo {author} {\bibfnamefont {A.}~\bibnamefont {Singh}},\ }\href@noop {} {\bibfield  {journal} {\bibinfo  {journal} {Journal of Theoretical Biology}\ }\textbf {\bibinfo {volume} {598}},\ \bibinfo {pages} {111996} (\bibinfo {year} {2025})}\BibitemShut {NoStop}%
\bibitem [{\citenamefont {Tang}\ \emph {et~al.}(2024)\citenamefont {Tang}, \citenamefont {Liu}, \citenamefont {Zhu}, \citenamefont {Cheng}, \citenamefont {Liu}, \citenamefont {Hammerschmidt}, \citenamefont {Zhou},\ and\ \citenamefont {Cai}}]{tang2024bet}%
  \BibitemOpen
  \bibfield  {author} {\bibinfo {author} {\bibfnamefont {S.}~\bibnamefont {Tang}}, \bibinfo {author} {\bibfnamefont {Y.}~\bibnamefont {Liu}}, \bibinfo {author} {\bibfnamefont {J.}~\bibnamefont {Zhu}}, \bibinfo {author} {\bibfnamefont {X.}~\bibnamefont {Cheng}}, \bibinfo {author} {\bibfnamefont {L.}~\bibnamefont {Liu}}, \bibinfo {author} {\bibfnamefont {K.}~\bibnamefont {Hammerschmidt}}, \bibinfo {author} {\bibfnamefont {J.}~\bibnamefont {Zhou}},\ and\ \bibinfo {author} {\bibfnamefont {Z.}~\bibnamefont {Cai}},\ }\href@noop {} {\bibfield  {journal} {\bibinfo  {journal} {Nature Communications}\ }\textbf {\bibinfo {volume} {15}},\ \bibinfo {pages} {2063} (\bibinfo {year} {2024})}\BibitemShut {NoStop}%
\bibitem [{\citenamefont {Gardiner}(2009)}]{gardiner2009stochastic}%
  \BibitemOpen
  \bibfield  {author} {\bibinfo {author} {\bibfnamefont {C.}~\bibnamefont {Gardiner}},\ }\href@noop {} {\emph {\bibinfo {title} {Stochastic Methods}}},\ Vol.~\bibinfo {volume} {4}\ (\bibinfo  {publisher} {Springer Berlin},\ \bibinfo {year} {2009})\BibitemShut {NoStop}%
\bibitem [{\citenamefont {Biswas}\ \emph {et~al.}(2021)\citenamefont {Biswas}, \citenamefont {Manicka}, \citenamefont {Hoel},\ and\ \citenamefont {Levin}}]{biswas2021gene}%
  \BibitemOpen
  \bibfield  {author} {\bibinfo {author} {\bibfnamefont {S.}~\bibnamefont {Biswas}}, \bibinfo {author} {\bibfnamefont {S.}~\bibnamefont {Manicka}}, \bibinfo {author} {\bibfnamefont {E.}~\bibnamefont {Hoel}},\ and\ \bibinfo {author} {\bibfnamefont {M.}~\bibnamefont {Levin}},\ }\href@noop {} {\bibfield  {journal} {\bibinfo  {journal} {Iscience}\ }\textbf {\bibinfo {volume} {24}} (\bibinfo {year} {2021})}\BibitemShut {NoStop}%
\bibitem [{\citenamefont {Lynch}\ and\ \citenamefont {Marinov}(2015)}]{lynch2015bioenergetic}%
  \BibitemOpen
  \bibfield  {author} {\bibinfo {author} {\bibfnamefont {M.}~\bibnamefont {Lynch}}\ and\ \bibinfo {author} {\bibfnamefont {G.~K.}\ \bibnamefont {Marinov}},\ }\href@noop {} {\bibfield  {journal} {\bibinfo  {journal} {Proceedings of the National Academy of Sciences}\ }\textbf {\bibinfo {volume} {112}},\ \bibinfo {pages} {15690} (\bibinfo {year} {2015})}\BibitemShut {NoStop}%
\bibitem [{\citenamefont {Mehta}\ \emph {et~al.}(2016)\citenamefont {Mehta}, \citenamefont {Lang},\ and\ \citenamefont {Schwab}}]{mehta2016landauer}%
  \BibitemOpen
  \bibfield  {author} {\bibinfo {author} {\bibfnamefont {P.}~\bibnamefont {Mehta}}, \bibinfo {author} {\bibfnamefont {A.~H.}\ \bibnamefont {Lang}},\ and\ \bibinfo {author} {\bibfnamefont {D.~J.}\ \bibnamefont {Schwab}},\ }\href@noop {} {\bibfield  {journal} {\bibinfo  {journal} {Journal of Statistical Physics}\ }\textbf {\bibinfo {volume} {162}},\ \bibinfo {pages} {1153} (\bibinfo {year} {2016})}\BibitemShut {NoStop}%
\bibitem [{\citenamefont {Cohen}(1966)}]{cohen1966optimizing}%
  \BibitemOpen
  \bibfield  {author} {\bibinfo {author} {\bibfnamefont {D.}~\bibnamefont {Cohen}},\ }\href@noop {} {\bibfield  {journal} {\bibinfo  {journal} {Journal of Theoretical Biology}\ }\textbf {\bibinfo {volume} {12}},\ \bibinfo {pages} {119} (\bibinfo {year} {1966})}\BibitemShut {NoStop}%
\bibitem [{\citenamefont {Seoane}\ and\ \citenamefont {Sol{\'e}}(2018)}]{seoane2018information}%
  \BibitemOpen
  \bibfield  {author} {\bibinfo {author} {\bibfnamefont {L.~F.}\ \bibnamefont {Seoane}}\ and\ \bibinfo {author} {\bibfnamefont {R.~V.}\ \bibnamefont {Sol{\'e}}},\ }\href@noop {} {\bibfield  {journal} {\bibinfo  {journal} {Royal Society Open Science}\ }\textbf {\bibinfo {volume} {5}},\ \bibinfo {pages} {172221} (\bibinfo {year} {2018})}\BibitemShut {NoStop}%
\bibitem [{\citenamefont {Cheong}\ \emph {et~al.}(2011)\citenamefont {Cheong}, \citenamefont {Rhee}, \citenamefont {Wang}, \citenamefont {Nemenman},\ and\ \citenamefont {Levchenko}}]{cheong2011information}%
  \BibitemOpen
  \bibfield  {author} {\bibinfo {author} {\bibfnamefont {R.}~\bibnamefont {Cheong}}, \bibinfo {author} {\bibfnamefont {A.}~\bibnamefont {Rhee}}, \bibinfo {author} {\bibfnamefont {C.~J.}\ \bibnamefont {Wang}}, \bibinfo {author} {\bibfnamefont {I.}~\bibnamefont {Nemenman}},\ and\ \bibinfo {author} {\bibfnamefont {A.}~\bibnamefont {Levchenko}},\ }\href@noop {} {\bibfield  {journal} {\bibinfo  {journal} {Science}\ }\textbf {\bibinfo {volume} {334}},\ \bibinfo {pages} {354} (\bibinfo {year} {2011})}\BibitemShut {NoStop}%
\bibitem [{\citenamefont {Hahn}\ \emph {et~al.}(2023)\citenamefont {Hahn}, \citenamefont {Walczak},\ and\ \citenamefont {Mora}}]{hahn2023dynamical}%
  \BibitemOpen
  \bibfield  {author} {\bibinfo {author} {\bibfnamefont {L.}~\bibnamefont {Hahn}}, \bibinfo {author} {\bibfnamefont {A.~M.}\ \bibnamefont {Walczak}},\ and\ \bibinfo {author} {\bibfnamefont {T.}~\bibnamefont {Mora}},\ }\href@noop {} {\bibfield  {journal} {\bibinfo  {journal} {Physical Review Letters}\ }\textbf {\bibinfo {volume} {131}},\ \bibinfo {pages} {128401} (\bibinfo {year} {2023})}\BibitemShut {NoStop}%
\end{thebibliography}%

\clearpage
\onecolumngrid

\makeatother

{\Large\textbf{Supplementary Information}}
\renewcommand{\thesection}{S\arabic{section}}

\setcounter{section}{0}  
\renewcommand{\thefigure}{S\arabic{figure}}

\setcounter{equation}{0} 
\renewcommand{\theequation}{S\arabic{equation}}
\section{Derivation of Fraction Rate Equation} \label{Appendix1}
Let $N_{y_j}$ be the number of individuals of phenotype $y_j$ in environment $x_i$,  $\in\mathbb{R}^+$; $i,j\in\{1,2\}$.~$N_{y_j}$ is growing exponentially at rate $g_{y_1}^{x_i}$.~Assuming that $N_{y_j}$ is high enough, its dynamics can be described by continuous differential equations in a given environmental state $x_i$ as follows:
\begin{subequations}
\begin{align}
    \frac{dN_{y_1}}{dt}&= g_{y_1}^{x_i}N_{y_1}-k_{y_1y_2}^{x_i}N_{y_1}+k_{y_2y_1}^{x_i}N_{y_2}, \label{eq1}\\
    \frac{dN_{y_2}}{dt}&= g_{y_2}^{x_i}N_{y_2}+k_{y_1y_2}^{x_i}N_{y_1}-k_{y_2y_1}^{x_i}N_{y_2}, \\
    \frac{d{(N_{y_1}+N_{y_2})}}{dt}&= g_{y_1}^{x_i}N_{y_1}+g_{y_2}^{x_i}N_{y_2}.~\label{eq2}
\end{align}
\end{subequations}
In a fixed environment \( x_i \), we define \( f_{y_j} \) as the fraction of individuals exhibiting phenotype \( y_j \) relative to the total population size.~Specifically, this implies
$f_{y_1} = \frac{N_{y_1}}{N_{y_1} + N_{y_2}}$.~We proceed to derive an explicit expression for the time derivative
$\frac{d}{dt} \left( \frac{N_{y_1}}{N_{y_1} + N_{y_2}} \right)$.

\begin{align}
    \frac{df_{y_j}}{dt}=\frac{d}{dt}\left(\frac{N_{y_1}}{N_{y_1}+N_{y_2}}\right)=&\frac{1}{N_{y_1}+N_{y_2}}\frac{dN_{y_1}}{dt} -\frac{N_{y_1}}{(N_{y_1}+N_{y_2})^2}\frac{d{(N_{y_1}+N_{y_2})}}{dt}
\end{align}
Using the system of equations \eqref{eq1} for the first term and \eqref{eq2} for the second term, we get:
\begin{align}
    \frac{d}{dt}\left(\frac{N_{y_1}}{N_{y_1}+N_{y_2}}\right)=&\frac{ ( g_{y_1}^{x_i}N_{y_1}-k_{y_1y_2}^{x_i}N_{y_1}+k_{y_2y_1}^{x_i}N_{y_2} )}{N_{y_1}+N_{y_2}} -\frac{N_{y_1}(g_{y_1}^{x_i}N_{y_1}+g_{y_2}^{x_i}N_{y_2})}{(N_{y_1}+N_{y_2})^2}
\end{align}
Using the fraction expressions of $f_{y_2}=1-f_{y_1}=\frac{N_{y_2}}{N_{y_1}+N_{y_2}}$ and $f_{y_1}=\frac{N_{y_1}}{N_{y_1}+N_{y_2}}$, and simplifying, we obtain,
\begin{align}\label{generalized_eq}
    \frac{df_{y_1}}{dt}=k^{x_i}_{y_2y_1}(1-f_{y_1}) - k^{x_i}_{y_1y_2} f_{y_1}  +(g^{x_i}_{y_1} - g^{x_i}_{y_2})  f_{y_1} (1 - f_{y_1})
\end{align}
Similarly, calculating for the phenotype $y_2$, we will get the generalized equation in \eqref{general_model}.

\section{Derivation of Numerator of $\Gamma$ }\label{Appendix2}
To find the numerator of $\Gamma$ which is $\gamma-\langle\gamma\rangle_{ind}$, we used the equations \eqref{gamma_prob} and \eqref{gamma_id}, and obtain the following,

\begin{align}
\langle\gamma\rangle-\langle\gamma\rangle_{ind}=& g^{x_1}_{y_1} P_{x_1y_1} +g^{x_1}_{y_2}P_{{x_1}y_2}
     +g^{x_2}_{y_1} P_{{x_2}y_1} + g^{x_2}_{y_2} P_{{x_2}y_2} - g^{x_1}_{y_1} P_{x_1} P_{y_1} -g^{x_1}_{y_2}P_{x_1}P_{y_2} -g^{x_2}_{y_1} P_{x_2}P_{y_1} - g^{x_2}_{y_2} P_{x_2}P_{y_2} 
     \end{align}
After using the relation $P_{x_i}=\sum_{j=1}^{2}P_{x_iy_j}$, the expression becomes,
\begin{align}
     \langle\gamma\rangle-\langle\gamma\rangle_{ind}=& g^{x_1}_{y_1}(P_{x_1y_1}- P_{x_1} P_{y_1}) + g^{x_1}_{y_2}((P_{x_1}-P_{x_1y_1})-P_{x_1}P_{y_2}) +g^{x_2}_{y_1}((P_{x_2}-P_{x_2y_2})-P_{x_2}P_{y_1})+g_{y_2}^{x_2} (P_{x_2y_2}-P_{x_2}P_{y_2})
\end{align}
By rearranging in terms of probability, we can effectively express the differences in growth rates,
\begin{align}
    \langle\gamma\rangle-\langle\gamma\rangle_{ind}=& (g^{x_1}_{y_1}-g^{x_1}_{y_2}) P_{x_1y_1}+(g^{x_2}_{y_2}-g^{x_2}_{y_1})P_{x_2y_2} - g^{x_1}_{y_1}P_{x_1} P_{y_1} + g^{x_1}_{y_2}(P_{x_1} -P_{x_1}P_{y_2})
         -g^{x_2}_{y_2}P_{x_2}P_{y_2}+g^{x_2}_{y_1}(P_{x_2}-P_{x_2}P_{y_1})
\end{align}
Using $g_{1}=g_{y_1}^{x_1}-g_{y_2}^{x_1}$ and $g_{2}=g_{y_2}^{x_2}-g_{y_1}^{x_2}$ the expression of relative growth rates and using the relation $P_{y_j}=\sum_{i=1}^{2}P_{x_i y_j}$ we get,
\begin{align}
    \langle\gamma\rangle-\langle\gamma\rangle_{ind}=&(g^{x_1}_{y_1}-g^{x_1}_{y_2})P_{x_1y_1}+(g^{x_2}_{y_2}-g^{x_2}_{y_1})P_{x_2y_2}-g^{x_1}_{y_1}P_{x_1} P_{y_1}+g^{x_1}_{y_2}P_{x_1}P_{y_1}
    -g^{x_2}_{y_2}P_{x_2}P_{y_2} +g^{x_2}_{y_1}P_{x_2}P_{y_2}
\end{align}
Simplifying and using the relative growth equation, we get
\begin{align}
    \langle\gamma\rangle-\langle\gamma\rangle_{ind}=&(g^{x_1}_{y_1}-g^{x_1}_{y_2})P_{x_1y_1}+(g^{x_2}_{y_2}-g^{x_2}_{y_1})P_{x_2y_2}-(g^{x_1}_{y_1}-g^{x_1}_{y_2})P_{x_1} P_{y_1} -(g^{x_2}_{y_2}-g^{x_2}_{y_1})P_{x_2}P_{y_2}
\end{align}
Further simplification gives us,
\begin{align}
    \langle\gamma\rangle-\langle\gamma\rangle_{ind}=&g_{x_1}(P_{x_1y_1}-P_{x_1} P_{y_1})+g_{x_2}(P_{x_2y_2}-P_{x_2}P_{y_2})
\end{align}

\section{Analytical Derivation of the Expression for Fraction and Integration of the phenotype fraction}\label{Appendix3}
For the system mentioned in subsection \ref{persistent}, using the growth rates and switching rates, the system equation becomes

\begin{align}\label{persister model}
\frac{df_{y}}{dt} =
   k (1 - f_{y}) - k f_{y} +g f_{y} (1 - f_{y}) 
\end{align}
The fraction $f_{y}$ denotes the fraction of the fittest phenotype in a given environment.~A phenotype is considered "fittest" when it exhibits the highest growth rate among the two phenotypes, $y_1$ and $y_2$, in their corresponding environments, $x_1$ or $x_2$.~So $g$ is positive when the phenotype aligns with their optimal environment and negative when it does not.~To proceed with solving the equation, we observe that the right-hand side of the equation is a quadratic expression.~This allows us to rewrite it as the product of two factors based on its roots.~Taking the root as $f_1$ and $f_2$, we can express the equation in terms of these roots:
\begin{align}
    f_1=\frac{g-2k-\sqrt{(g^2+4k^2)}}{2g},\\
    f_2=\frac{g-2k+\sqrt{(g^2+4k^2)}}{2g}
\end{align}
Using this and rearranging \eqref{persister model}, we obtain:
\begin{align}
\frac{df_{y}}{dt} =g(f_{y}-f_1)(f_{y}-f_2)
\end{align}
We analytically derive the expression for the fraction of the population considering a time interval of duration
\( \Delta t_i \).~Here, time \( t = 0 \) denotes the moment the system enters a specific environment, and time
\( t = \Delta t_i \) corresponds to the instant just before the system leaves that environment.~
The resulting expression characterizes the fraction of the population after evolving in a particular environment for a period of time
\( \Delta t_i \).

\begin{align}
    \int_{0}^{\Delta t_i} \frac{1}{(f_{y} - f_1)(f_{y} - f_2)} \, df_{y_j} 
    &= \int_{0}^{\Delta t_i} (g) \, dt \notag \\
    \Rightarrow \int_{0}^{\Delta t_i} \frac{1}{f_1 - f_2}
    \left( \frac{1}{f_{y} - f_1} - \frac{1}{f_{y} - f_2} \right) df_{y} 
    &= \int_{0}^{\Delta t_i} (g) \, dt \notag \\
    \Rightarrow \frac{1}{f_1 - f_2}
    \ln \left( \frac{(f_{y}(\Delta t_i) - f_1)(f_{y}(0) - f_2)}
                     {(f_{y}(\Delta t_i) - f_2)(f_{y}(0) - f_1)} \right) 
    &= g \Delta t_i
\end{align}
Solving for $f_{y}$, the expression for the fraction can be derived in the following form:
\begin{align}\label{fraction}
   f_{y}(\Delta t_i) = \frac{a+ce^{-b\Delta t_i}}{d+he^{-b\Delta t_i}}
\end{align}
where
\begin{subequations}\label{expression}
\begin{eqnarray}
    a &=& 2k-f_{y}(0)(g+2k)+f_{y}(0)\sqrt{g^2+4k^2}\\
    b &=& \sqrt{g^2+4k^2}\\
    c &=& f_{y}(0)(g+2k)-2k+\sqrt{g^2+4k^2}\\
    d &=& (g+2k)-2f_{y}(0)g+\sqrt{g^2+4k^2}\\
    h &=& 2f_{y}(0)g-(g+2k)+\sqrt{g^2+4k^2}
\end{eqnarray}
\end{subequations}
The initial condition for $f_{y}(t)$ is denoted by $f_y(0)$.~To compute the integral of \( f_{y}(\Delta t_i) \), we calculate the area under the curve \( f_{y}(\cdot) \) over the time interval \( \Delta t_i \),

\begin{align}
    \int_{0}^{\Delta t_i} f_{y}(t)\, dt 
    &= \frac{abh\Delta t_i}{bdh}
     - \frac{(cd - ah)}{bdh} 
    \Big( \ln(d + h)  - \ln(d + h e^{-b\Delta t_i}) \Big)\notag\\
\end{align}
For simplicity, we denote this integral as \( \mathcal{I}_{y}(\Delta t_i) \).~Substituting the expressions in eq.\eqref{expression} for the parameters, we obtain the following,

\begin{align}\label{integration_fraction}
    \mathcal{I}_{y}(\Delta t_i)&=\int_{0}^{\Delta t_i} f_{y}(t)\, dt 
    = \frac{(g - 2k + \sqrt{g^2 + 4k^2}) \Delta t_i}{2g}  + \frac{1}{g} \ln \Bigg( 
        \frac{(2f_{y}(0) + 2k - g)(1 - e^{-\sqrt{g^2 + 4k^2}\Delta t_i})}
        {2\sqrt{g^2 + 4k^2}} + \frac{1 + e^{-\sqrt{g^2 + 4k^2}\Delta t_i}}{2} 
    \Bigg)
\end{align}
This integral is used to calculate the time-averaged statistics of the joint distributions between phenotypes and environments.~
The total integration for the fittest phenotype $y$ is given by 
\begin{equation}
    \mathcal{I}_{y}(t)=\sum_{i=1}^n \mathcal{I}_{y}(\Delta t_i)
\end{equation}
where each $\Delta t_i$ represents a time interval.~This cumulative measure can then be used to derive the corresponding probability~distribution.

\section{Simulation For Fig.\ref{fig:persistence}}

This algorithm calculates the evolution of the population fraction and therefore the information and fitness of a population of individuals expressing either two phenotypic states, subjected to an environment that fluctuates randomly between a stress condition and a normal growth condition.~Specifically, our system considers and environment environmental switching rates $\lambda_{12}$ and $\lambda_{21}$ and therefore the environmental distribution $P_{x_1}$ and ($P_{x_2}=1-P_{x_1}$) and phenotypes with distribution \(P_{y_1}\).~Given the relative growth rates $\mu_1$ and $\mu_2$, we aim to calculate the average population fitness \(\langle \gamma \rangle\), the independent population fitness \(\langle \gamma \rangle_{\mathrm{ind}}\), the normalized population fitness \(\Gamma\), and the mutual information \(I\).

The simulation is performed for given values of $P_{x_1}$, $\lambda_{21}$ and $\mu_1$.~For each pair $P_{x_1}$, $\lambda_{21}$, the corresponding environmental switching rate from stress ($x_1$) to normal ($x_2$), denoted $\lambda_{12}$, is calculated.~The simulation begins by initializing the population’s phenotype as the sensitive type $y_1$, with its fraction $f_s$  drawn from an uniform distribution between $0$ and $1$.~The environment is then randomly set to either the proliferating or stress state.

The simulation then proceeds for \(N\) iterations or environmental transitions.~At each iteration, the time until the environment switches again is sampled from an exponential distribution parameterized by the current environment’s switching rate (e.g.~$\lambda_{12}$).~If the current environment is stressful, the algorithm evaluates how much time within this interval the population spends in each phenotype state by applying an indicator function and updates the counters accordingly.~

When the system is in a stress environment, the algorithm updates the phenotypic distribution using a phenotype evolution function, denoted by \( f_{y_1}(\Delta t; \mu_1) \), which follows the solution \eqref{fraction}.~In this context, the dormant phenotype, $y_1$ is the fittest for the environment $x_1$, and therefore $g$ is assigned a positive value.~This function describes how the fraction of individuals in the phenotype $y_1$ evolves over a time interval \( \tau \), given the initial condition \( f_s \), the death rate \( \mu_1 \) in the stress environment, and the switching rate, \( k \) between phenotypes.~In addition to this point-wise update, the algorithm calculates the integral \( \mathcal{I}_{y_1}(\Delta t;\mu_1) \), which represents the integration of the fraction of the phenotype $y_1$ during the time spent in the stress environment \eqref{integration_fraction}.~After these updates are performed, the environment switches from $x_1$ to $x_2$, and a similar set of operations is carried out using the corresponding function and integral for the normal environment.

A similar series of procedures is followed in the $x_1$ environment: the time spent in that environment is sampled, and the fractions and integrals are updated.~The environment then switches back to the stress state.~At each iteration, the total simulation time is incremented by the duration \( \Delta t \) of the current environmental state.

After completing all iterations, the algorithm normalizes the time duration of each environmental phenotype by the total simulation time to obtain joint probability distributions of phenotype-environment pairs \(P_{x_iy_j}\).~Marginal probabilities for each phenotype \(P_{y_j}\) and each environment state probabilities \(P_{x_i}\) are then calculated by summing over appropriate joint probabilities, $\hat{\mathcal{I}}^t_{x_1,y_1}$.~These probabilities form the basis for computing key quantities: the average population fitness \( \langle \gamma\rangle\), the independent population fitness \(\langle\gamma\rangle_{ind}\), the normalized population fitness \(\Gamma\), and environmental states; and the mutual information \(I\) as explained in the main text.

Finally, the collected data enable a comprehensive analysis of how dormant phenotype switching, environmental stochasticity, and stress-induced death rates interact to shape various population fitness measures and mutual information, as illustrated in the accompanying plots Fig.~\ref{fig:persistence}.

\begin{algorithm}[H]
\caption{This pseudo-code designed to generate the diagram in Fig \ref{fig:persistence} for fixed $P_{x_1}$, $\lambda_{21}$ and $\mu_1$.~All formulas and notations are used consistently and follow standard conventions throughout the manuscript.}
\KwIn{\\
    $\lambda_{21}$: rate of switching from $x_1$ and $x_2$ (normal to antibiotic environment) \\
    $\mu_2$: growth rate in non-stressful environment \\
    $k$: switching rate between phenotypes \\
    $P_{x_1}$: Stressful environment probability \\
    $\mu_1$: Death rate in stress environment \\
    $N$: number of iterations per simulation
}
\KwOut{\\
Simulation statistics including:\\
$P_{x_1}, P_{x_2}, P_{y_1}, P_{y_2}, P_{x_1 y_1}, P_{x_1 y_2}, P_{x_2 y_1}, P_{x_2 y_2}, P, \langle \gamma \rangle_{ind}, \langle \gamma \rangle, \Gamma, I$}

Compute $\lambda_{12} = \lambda_{21} \cdot \frac{1 - P_{x_1}}{P_{x_1}}$ \\
Compute $P_2 = \lambda_{12} / (\lambda_{12} + \lambda_{21})$ \\

Initialize phenotype fraction $f_s \sim \mathcal{U}(0,1)$ \\
Initialize $\texttt{env} \sim \text{Uniform}\{x_1, x_2\}$  \\
Set total time $t = 0$ \\
Initialize counters: $\hat{P}_{11}^t, \hat{P}_{12}^t, \hat{P}_{21}^t, \hat{P}_{22}^t = 0$ \\

\For{$i = 1$ \KwTo $N$}{
    \eIf{$env = 1$}{
        Sample $\Delta t_i \sim \text{Exponential}(1 / \lambda_{12})$ \\
        $\hat{Z}_{11} += \mathcal{{I}}_{y_2}(\Delta t_i;\mu_1)$ \\
        $\hat{Z}_{12} += \Delta t_i - {\mathcal{I}}_{y_2}(\Delta t; \mu_1)$ \\
        $f_s = f_{y_2}(\Delta t_i; \mu_1)$ \\
        Switch to $env = 2$
    }{
        Sample $\Delta t_i \sim \text{Exponential}(1 / \lambda_{21})$ \\
        $\hat{Z}_{21} += {\mathcal{I}}_{y_2}(\Delta t_i; -\mu_2)$ \\
        $\hat{Z}_{22} += \Delta t_i - {\mathcal{I}}_{y_2}(\Delta t_i; -\mu_2)$ \\
        $f_s = f_{y_2}(\Delta t_i; -\mu_2)$ \\
        Switch to $env = 1$
    }
    $t += \Delta t$
}

Distributions:
\[
P_{x_iy_j} = \frac{\hat{Z}_{ij}^t}{t}, \quad \text{for } i,j \in \{1,2\}
\]

Marginal Distributions:
\[
P_{x_1} = P_{x_1y_1} + P_{x_1y_2}, \quad P_{x_2} = P_{x_2y_1} + P_{x_2y_2}
\]
\[
P_{y_1} = P_{x_1y_1} + P_{x_2y_1}, \quad P_{y_2} = P_{x_1y_2} + P_{x_2y_2}
\]

Growth Rates and Information:
\[
\gamma = -\mu_1 \cdot P_{x_1y_2} + \mu_2 \cdot P_{x_2y_2}
\]
\[
\gamma_{{ind}} = -\mu_1 \cdot P_{x_1} \cdot P_{y_2} + \mu_2 \cdot P_{x_2} \cdot P_{y_2}
\]
\[
\Gamma = \frac{\mu_1 (P_{x_1y_1} - P_{x_1} P_{y_1}) + \mu_2 (P_{x_2y_2} - P_{x_2} P_{y_2})}{(\mu_1 + \mu_2) P_{x_1} P_{x_2}}
\]
\[
I = \sum_{i,j} P_{x_i y_j} \log_2 \left( \frac{P_{x_i y_j}}{P_{x_i} \cdot P_{y_j}} \right)
\]

Return all computed quantities in a data structure.
\end{algorithm}

\section{Simulation For Fig.\ref{fig:switching_fraction_compact}B}

This algorithm, used for plotting Fig.~\ref{fig:switching_fraction_compact}B,  is used for plotting the trajectories of the mean fraction in the fluctuating environment.~The system transitions between two environments: $x_1$ (stress environment) and $x_2$ (growth environment), with switching rates $\lambda_{12}$ and $\lambda_{21}$, respectively.~The slow-growing phenotypic fraction $f_s$ (corresponding to $y_1$) evolves over discrete time steps $\Delta t_i$ according to the analytical expression $f_{y_1}(t)$, which depends on the death or growth rate ($\mu_1$ in $x_1$, $-\mu_2$ in $x_2$) and the phenotypic switching rate $k$.~At each step, the environment changes with probability proportional to the product of the rate of the environmental switch $\lambda_{ij}$ and the time step.~The current time $t$, the fraction of phenotypes $f_s$, and the state of the environment $\texttt{env}$ are stored in an array, which is later converted into a DataFrame for plotting.~The resulting trajectory illustrates how the phenotypic composition dynamically responds to stochastic environmental changes, as shown in Fig.~\ref{fig:switching_fraction_compact}B.

\begin{algorithm}[H]
\caption{Simulate time evolution of phenotype fraction and environment to generate trajectory plot of Fig~\ref{fig:switching_fraction_compact}B.}
\KwIn{\\
    $\lambda_{12}$: switch rate from environment $x_1$ (antibiotic) to $x_2$ (normal) \\
    $\lambda_{21}$: switch rate from environment $x_2$ to $x_1$ \\
    $\mu_1, \mu_2$: growth/death rates in environments $x_1$, $x_2$ \\
    $k$: phenotype switching rate \\
    $f_s$: initial phenotype fraction \\
    $\Delta t_i$: fixed time step \\
    $N$: total number of iterations
}
\KwOut{\\
Time series of $f_s(t)$ and environment $\texttt{env}$ used for plotting trajectories}

Initialize: $f_s \sim \mathcal{U}(0,1)$, $t = 0$, $\texttt{env} \sim \text{Uniform}\{x_1, x_2\}$ \\
Initialize empty array \texttt{dataarr} = [\ ] \\

\For{$i = 1$ \KwTo $N$}{
    \eIf{$\texttt{env} = x_1$ }{
        Update phenotype: $f_s \leftarrow f_{y_1}(\Delta t_i; \mu_1)$ \\
        With probability $\lambda_{12} \cdot \Delta t_i$, set $\texttt{env} \leftarrow x_2$
    }{
        Update phenotype: $f_s \leftarrow f_{y_1}(\Delta t_i; -\mu_2)$ \\
        With probability $\lambda_{21} \cdot \Delta t_i$, set $\texttt{env} \leftarrow x_1$
    }
    $t \leftarrow t + \Delta t_i$ \\
    Append $[t, f_s, \texttt{env}]$ to \texttt{dataarr}
}

Create DataFrame $D$ from \texttt{dataarr} with columns: \texttt{Time}, \texttt{Fraction}, \texttt{Env} \\


\Return{Trajectory plot showing $f_{y_1}(t)$ dyanamics with environmental switching}
\end{algorithm}

\end{document}